\documentclass[journal]{IEEEtran}

\usepackage{amssymb}
\usepackage{amsmath}
\usepackage{makeidx}
\usepackage{multicol}
\usepackage[abs]{overpic}
\usepackage{tabularx}
\usepackage[numbers,sort&compress]{natbib}
\usepackage{amsfonts}
\usepackage{color}
\usepackage{array,color}
\usepackage{tabularx}
\usepackage{multirow}
\usepackage{graphics}
\usepackage{epsfig}
\usepackage{graphicx}
\usepackage{makeidx}
\usepackage{multicol}
\usepackage{amsthm}
\usepackage{bm}
\usepackage{booktabs}
\usepackage{ulem}
\usepackage{enumerate}
\usepackage{subfig}
\usepackage{epstopdf}

\newtheorem{theorem}{theorem}[section]

\newtheorem{proposition}[theorem]{Proposition}

\begin{document}
%
\title{Massive Streaming PMU Data Modeling and Analytics in Smart Grid State Evaluation Based on Multiple High-Dimensional Covariance Tests}


\author{Lei Chu,
        Robert Qiu,~\IEEEmembership{Fellow,~IEEE}, Xing He, Zenan Ling and Yadong Liu
\IEEEcompsocitemizethanks{\IEEEcompsocthanksitem Dr. Qiu is with the Department of Electrical and Computer Engineering\, Tennessee Technological University, Cookeville, TN 38505 USA. Dr. Qiu is also with Department of Electrical Engineering, Research Center for Big Data Engineering Technology, State Energy Smart Grid Resarch and Development Center, Shanghai Jiaotong University, Shanghai 200240, China. (e-mail: rcqiu@sjtu.edu.cn; rqiu@tntech.edu). \protect
\IEEEcompsocthanksitem Lei Chu, Zenan Ling, Yadong Liu and Xing He are with Department of Electrical Engineering, Research Center for Big Data Engineering Technology, State Energy Smart Grid Resarch and Development Center, Shanghai Jiaotong University, Shanghai 200240, China. (Email: Leochu, zenanling and yadongliu@sjtu.edu.cn; hexing\_hx@126.com)}
\thanks{Dr.  Qiu's  work  is  supported  by N.S.F. of  China  No.61571296 and N.S.F.  of  US  Grant No.  CNS-1247778, No.  CNS-1619250. }}



\IEEEtitleabstractindextext{%
\begin{abstract}
The analogous deployment of phase measurement units (PMUs), the increase of data quantum and the deregulation of energy market, all call for the robust state evaluation in large scale power systems. Implementing model based estimators is impractical because of the complexity scale of solving the high dimension power flow equations. In this paper, we first represent massive streaming PMU data as big random matrix flow. By exploiting the variations in the covariance matrix of the massive streaming PMU data, a novel power state evaluation algorithm is then developed based on the multiple high dimensional covariance matrix tests. The proposed test statistic is flexible and nonparametric, which assumes no specific parameter distribution or dimension structure for the PMU data. Besides, it can jointly reveal the relative magnitude, duration and location of a system event. For the sake of practical application, we reduce the computation of the proposed test statistic from $O(\varepsilon n_g^4)$ to $O(\eta n_g^2)$ by principal component calculation and redundant computation elimination. The novel algorithm is numerically evaluated utilizing the IEEE 30-, 118-bus system, a Polish 2383-bus system, and a real 34-PMU system. The case studies illustrate and verify the superiority of proposed state evaluation indicator.
\end{abstract}

}

\maketitle

\IEEEpeerreviewmaketitle
\IEEEdisplaynontitleabstractindextext

\section{Introduction}
\label{Sec:1}
\IEEEPARstart{R}{eliable} operation and intelligent management of electric power systems have a heavy influence on daily life. Recently, power companies, scholars and researchers keep an eye on utilizing PMUs to improve wide area monitoring, protection, and control (WAMPAC) \cite{terzija2011wide, qiu2015smart}. Some large-scale implementations of synchrophasor technology for managing the power grids across the world have been brought online. As an illustration, there were about 2400 PMUs deployed in power grids in China as of 2013 \cite{lu2015advancing}; North America and India have coverage from about 2000 and 1800 PMUs by 2015, respectively \cite{nuthalapati2015managing}. Accordingly, designing, monitoring, and controlling such systems are becoming increasingly more challenging as a consequence of the steady growth of their size, complexity, level of uncertainty, unpredictable behavior, and interactions \cite{chakrabarti2009measurements, de2010synchronized, Pan2016Analyzing}.

Efforts are in place to take synchrophasor technology to evaluate power states and develop reliable operational procedures to better understand and manage the power grids with wide-area visualization tools using PMU data. These power state evaluation methods can be generally organized into two categories: model-based estimators and data driven estimators. Model-based analysis is a kind of traditional method for offline analysis of state evaluation in power systems. Lof and Anderson presented statistical state evaluation indices based on the largest singular value of the inverse of the power flow Jacobian matrix \cite{lof1993voltage}. Ghiocel and Chow extended the result in \cite{lof1993voltage} and identified power flow control infeasibilities in a large-scale power system \cite{ghiocel2014power}. Pordanjani, Wang and Xu assessed the state evaluation using Channel components transform  \cite{pordanjani2013identification}. More recently, equivalent nodal analysis for power state evaluation was shown in \cite{lee2016voltage}. With the help of eigenvalues, eigenvectors, and participation factors of the power flow Jacobian matrix, the system characteristics can be predicted by these estimators. However, they hardly meet the severe requirements for efficient and stable monitoring of dynamically changed power systems possessing the steady growth of their size, complexity, and unpredictable behaviors.

As a novel alternative, the latest advanced data driven estimators can assess state evaluation without knowledge of the power network parameters or topology \cite{xu2015correlation, xie2014dimensionality, lim2016svd, ghanavati2016identifying, he2015big, hedesigning}. A linearized analysis algorithm was proposed for early event detection using the reduced dimensionality \cite{xie2014dimensionality}. Lim and DeMarco presented a SVD-based power state evaluation from PMU data, but their methods would be difficult to implement for real time assessment in a large power system due to the high computation burden \cite{lim2016svd}. Instead of monitoring the raw PMU data, recently, there has been considerable interest in the statistics of PMU measurements. Ghanavati, Hines, and Lakoba sought to identify a statistical state evaluation indicator by calculating the expected variance and autocorrelation of the buses' voltages and currents \cite{ghanavati2016identifying}. It is noted that the success of these approaches requires an accurate statistical model of measurement noise and load fluctuations. Besides, the constraint that the data dimension should be smaller than the window size also needs to be satisfied in \cite{xie2014dimensionality, lim2016svd}. On the other hand, linear eigenvalue statistics (LESs) of the high-dimensional PMU data were utilized for situational awareness or correction analysis of the power system in our recent works \cite{xu2015correlation, he2015big, hedesigning}. Taking advantage of asymptotic properties of high-dimensional random matrix, LES-based methods provided robust power state evaluation using individual window-truncated PMU data. Rather than exploiting individual window-truncated PMU data, this work tries to indicate state evaluation by high-dimensional statistical properties of overall PMU data.

Besides, from the perspective of theoretical research, large deployment of synchronized PMU raises several open issues:
\begin{enumerate}
\item How to represent the massive streaming PMU data in the manner of continuous learning of a power system;
\item How to evaluate the real time state evaluation from massive streaming PMU data;
\item Is there any method that can turn the big PMU data into tiny data for the practical use?
\item How to develop a state evaluation estimator without assuming a specific parametric distribution for the data;
\item Does a flexible data driven state evaluation indicator with a wide range of dimensions and sample size exist?
\end{enumerate}

The new metric proposed here is based on multiple high dimensional covariance matrix tests. Tests about high-dimensional covariance matrices have recently increased in popularity. The first attempt on the high-dimensional covariance matrix test presented by Bai and Saranadasa was based on likelihood radio (LR) test \cite{bai1996effect}. The LR test works well for normally distributed data on condition that the sample size is larger than the data dimension. Gupta and Xu extended the LR test to non-normal distribution \cite{gupta2006some} while Bai et al. \cite{bai2009corrections} considered a correction of the LR (CLR) test in the case of a wide range of data dimensions. These tests share the basic assumption that the population covariance matrix can be directly substituted by the sample covariance matrix. However, genomic studies showed that such an assumption may not work because these sample covariance matrix based estimators have unnecessary terms which slow down the convergence considerably as the dimension is high \cite{ledoit2002some, chen2010two, chen2012tests}. Instead of estimating the population covariance matrix directly, some well-defined distance were proposed to evaluate the difference among populations \cite{chen2012tests}. Ledoit and Wolf exploited scaled trace-based distance measure between two sub-populations when the data dimension is large compared to the sample size \cite{ledoit2002some}. By exploiting the merits of U-statistics \cite{lee1990u}, Chen etc. extended the results in \cite{ledoit2002some} to a wide range of data dimension and sample size. However, these works are of high computation burden and focus on the difference of two sub-populations which make them unsuitable for indicating real time state evaluation in massive streaming PMU data.

In this paper, by exploiting the changes in the covariance matrix of different sampling periods of the streaming PMU data, we develop a novel power state evaluation algorithm using the multiple high dimensional covariance matrix tests. The key features of the proposed test statistic are as follows. 1) it can jointly reveal the relative magnitude, duration (or so-called clearing time) and location of a system event; 2) it specifies no parameter distribution of the PMU data, which implies a wide range of the practical applications; 3) it is a real time data driven method without requiring any knowledge of the system model or topology; 4) it is a flexible state evaluation indicator without specifying an explicit relationship between data dimension and sample size; 5) it provides effective computation due to principal component calculation and redundant computation elimination. 6) it implements the asymptotic properties of the high dimensional PMU data to enhance the robustness of the test statistic.

The remainder of this paper is structured as follows. Section \ref{Sec:2} introduces the representation of massive streaming PMU data. Section \ref{Sec:3} presents a power state evaluation approach using multiple high dimension covariance matrix tests. By principal component calculation and redundant computation elimination, an effective calculating method for the proposed test statistic is also developed. In Section \ref{sec:case}, numerical case studies using synthetic data and real data are provided to evaluate the performance of the proposed state evaluation indicator; The discussion is also included in this section. The conclusion is presented in Section \ref{sec:Conclusion}. For the sake of simplicity, all technical details and some additional case study results are deferred to the Appendices.

\section{Massive Streaming PMU Data Modeling}
\label{Sec:2}

It is well accepted that the transient behavior of a large electric power system can be illustrated by a set of differential and algebraic equations (DAEs) as follows \cite{milano2013systematic, bollen2000understanding}:
\begin{eqnarray}
\label{eqC1}
{{{\bf{\dot x}}}^{\left( t \right)}} &=& f\left( {{{\bf{x}}^{\left( t \right)}},{{\bf{u}}^{\left( t \right)}},{{\bf{h}}^{\left( t \right)}},w} \right)\\
\label{eqC2}
0 &=& g\left( {{{\bf{x}}^{\left( t \right)}},{{\bf{u}}^{\left( t \right)}},{{\bf{h}}^{\left( t \right)}},w} \right)
\end{eqnarray}
where ${{\bf{x}}^{\left( t \right)}} \in {\mathcal{C}}^{m \times p}$ are the power state variables, e.g., rotor speeds and the dynamic states of loads, ${{\bf{u}}^{\left( t \right)}}$ represent the system input parameters, ${{\bf{h}}^{\left( t \right)}}$ are the algebraic variables, e.g., bus voltage magnitudes, $w$ denotes the time-invariant system parameters. $t \in \mathcal{R}$, $m$, and $p$ are the sample time, number of system variables, and bus, respectively. The model-based state indicators \cite{ghiocel2014power, milano2013systematic, dorfler2013kron, lee2016voltage, ghanavati2016identifying} focus on linearization of nonlinear DAEs in \eqref{eqC1} and \eqref{eqC2} which gives
\begin{equation}
\label{eqC3}
\left[ {\begin{array}{*{20}{c}}
{\Delta {\bf{\uline{\dot x}}}}\\
{\Delta {\bf{\uline{\dot u}}}}
\end{array}} \right]{\rm{ = }}\left[ {\begin{array}{*{20}{c}}
{\bf{A}}&{ - {{\bf{f}}_{\bf{u}}}{\bf{g}}_{\bf{u}}^{ - 1}{{\bf{g}}_{\bf{h}}}}\\
{\bf{0}}&{ - {\bf{E}}}
\end{array}} \right]\left[ {\begin{array}{*{20}{c}}
{\Delta {\bf{\uline{x}}}}\\
{\Delta {\bf{\uline{u}}}}
\end{array}} \right]{\rm{ + }}\left[ {\begin{array}{*{20}{c}}
\bf{0}\\
{\bf{C}}
\end{array}} \right]{\bf{\xi }},
\end{equation}
where ${{\bf{f}}_{\bf{x}}}$, ${{\bf{f}}_{\bf{u}}}$ are the Jacobian matrices of ${{\bf{f}}}$ with respect to ${\bf{\uline{x}}}, {\bf{\uline{u}}}$ and ${\bf{A}} = {{\bf{f}}_{\bf{x}}} - {{\bf{f}}_{\bf{u}}}{\bf{g}}_{\bf{u}}^{ - 1}{{\bf{g}}_{\bf{x}}}$. $\bf{E}$ is a diagonal matrix whose diagonal entries equal $t_{cor}^{-1}$ and $t_{cor}$ is the correction time of the load fluctuations. $\bf{C}$ denotes a diagonal matrix whose diagonal entries are nominal values of the corresponding active $(P)$ or reactive $(Q)$ of loads; $\bf{\xi}$ is assumed to be a vector of independent Gaussian random variables.

It is noted that estimating the system state by solving the equation \eqref{eqC3} is becoming increasingly more challenging \cite{chakrabarti2009measurements, qiu2015smart} as a consequence of the steady growth of the parameters, say, $t$, $p$ and $m$. Besides, the assumption that $\bf{\xi}$ follows Gaussian distribution would restrict the practical application.

As a novel alternative, the lately advanced data driven estimators \cite{xu2015correlation, xie2014dimensionality, lim2016svd, ghanavati2016identifying, he2015big, hedesigning} can assess state evaluation without knowledge of the power network parameters or topology. However, these estimators are based on the analysis of individual window-truncated PMU data. In this work, we seek to provide a method with the ability of continuous learning of power system from massive streaming PMU data.

\begin{figure}[htbp]
{
\includegraphics[width=0.475\textwidth]{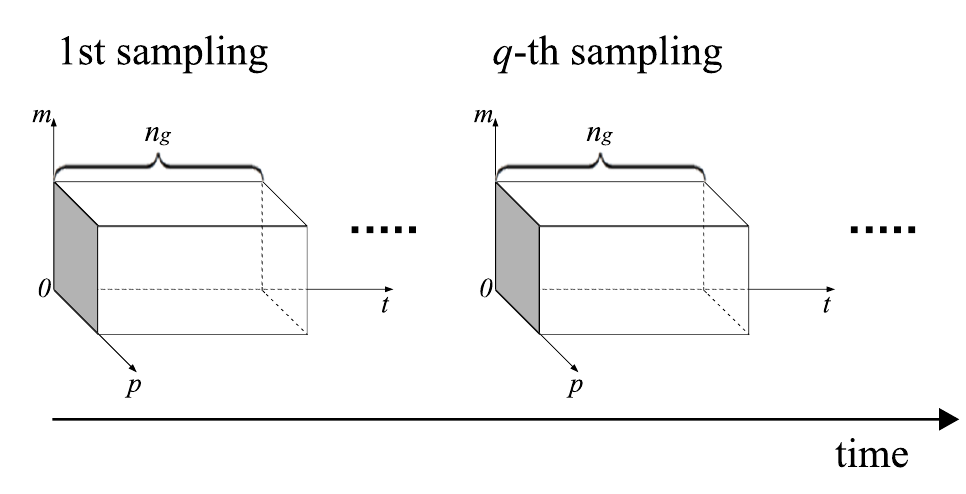}
}
\caption{{\label{fig1}} Conceptual representation of the structure of the massive streaming PMU data.}
\end{figure}

We first try to turn the big PMU data (massive high-dimensional PMU data streams)  into tiny data (PMU data segments) for  practical use. Fig. \ref{fig1} illustrates the conceptual representation of the structure of the massive streaming PMU data. More specifically, let $p$ denote the number of the available PMUs across the whole power network, each providing $m$ measurements. At $i$th time sample, a total of $\kappa = p \times m$ measurements, say ${\bf{z}}_i$, are collected. With respect to each PMU, the $m$ measurements could contain many categories of variables, such as voltage magnitude, power flow, and frequency, etc. In this work, we develop PMU data analysis assuming each type of measurements is independent. That is, we assume that at each round of analysis, $\kappa :=p$. Given $q$ time periods of $T$ seconds with $K$ Hz sampling frequency in $i$th data collection. Let $n_g=T \times K$ and $n$
be the window size and sampling number, respectively. A sequence of large random matrix
\begin{equation}
\label{eq1}
\left\{ {\underbrace {{{\bf{Z}}_{11}},{{\bf{Z}}_{12}}, \cdots ,{{\bf{Z}}_{1q}}}_{q \ {\rm{ window - truncated \ data}}}, \cdots ,\underbrace {{{\bf{Z}}_{n1}},{{\bf{Z}}_{n2}}, \cdots ,{{\bf{Z}}_{nq}}}_{q \ {\rm{ window - truncated \ data}}}} \right\}
\end{equation}
is obtained to represent the collected voltage magnitude data. Note that ${{\bf{Z}}_{ig}} = \left\{ {{{\bf{z}}_{i1}}, \cdots ,{{\bf{z}}_{{in_g}}}} \right\}, i = 1,2, \cdots, n$.

In this paper, magnitudes of bus voltages are employed as status data for the following considerations:
\begin{enumerate}
\item The voltage magnitudes are one of the most common measurements in power systems and plenty of studies are developed based on them.
\item The voltage magnitude can be collected with no prior knowledge of the topology of power systems. Therefore, we can conduct analysis without knowing the network structures and related parameters.
\item Previous studies \cite{xu2015correlation, hedesigning, xie2014dimensionality} have shown that power state indicators based on voltage magnitudes can achieve similar performances compared to other measurements, i.e., currents or power flow.
\end{enumerate}

\section{Voltage Stability Assessment}
\label{Sec:3}
Once we have established the random data flow model for the massive streaming PMU data, the next step is to extract the data analytics. As we all know, power systems are continuously experiencing fluctuations of small magnitudes \cite{bollen2000understanding}. In the functional setting, it is of interest to test whether or not $q$ sets of bus voltage curves have similar variations;  We assume that the system is initially in a steady-state operating condition for assessing the power state when subjected to a specified disturbance \cite{bollen2000understanding, kundur2004definition}. Thus it is interesting to discover the difference of the measurements collected in the normal condition and the abnormal condition by multiple high dimensional covariance matrix tests.

\subsection{Multiple High Dimensional Covariance Matrix Tests}

As depicted in the Section \ref{Sec:2}, a large random matrix flow $\left\{ {{{\bf{Z}}_1},{{\bf{Z}}_2}, \cdots ,{{\bf{Z}}_q}} \right\}$ is adopted to represent the massive streaming PMU data in one sample period. Instead of analyzing the raw individual window-truncated PMU data ${\left\{ {{{\bf{Z}}_g}} \right\}_{g = 1,2, \cdots q}}$ \cite{xie2014dimensionality, lim2016svd} or the statistic of ${{\bf{Z}}_g}$, \cite{xu2015correlation, ghanavati2016identifying, he2015big, hedesigning}, a comprehensive analysis of the statistic of $\left\{ {{{\bf{Z}}_1},{{\bf{Z}}_2}, \cdots ,{{\bf{Z}}_q}} \right\}$ is conducted in the following.

More specially, with ${{\bf{\Sigma }}_i}$ as the covariance matrix of $i$th collected PMU measurements, we test the hypothesis:
\begin{equation}
\label{eqB1}
\begin{array}{l}
{H_0}:{{\bf{\Sigma }}_1} = {{\bf{\Sigma }}_2} =  \cdots  = {{\bf{\Sigma }}_q} \\
{H_1}:\exists \ j,k \ \ {\rm{s.t.}} \ \ {{\bf{\Sigma }}_j} \ne {{\bf{\Sigma }}_k}
\end{array}.
\end{equation}

It is worthy noting that the hypothesis \eqref{eqB1} is a famous testing hypothesis in multivariate statistical analysis \cite[Chapter 10]{Anderson1959An} which aims to study samples that share or approximately share some of the same distributions and consider using a set of samples (data streams denote in equation (4) in this paper), one from each population, to test the hypothesis that the covariance matrices of these populations are equal.

\subsection{Proposed test statistic}
\label{ProsedTest}

The LR test \cite{bai1996effect} and CLR test \cite{bai2009corrections} as introduced in the Section \ref{Sec:1} are the most commonly test statistics for the hypothesis in \eqref{eqB1}. For the readers' convenience, we briefly explain the technical details in the Appendix A. These tests can be understood by replacing the population covariance matrix ${{\bf{\Sigma }}_g}$ by its sample covariance matrix ${\bf{Y}}_g$. Direct substitution of ${{\bf{\Sigma }}_g}$ by ${\bf{Y}}_g$ brings invariance and good testing properties as shown in \cite{bai1996effect} for normally distributed data. The test statistic $V_2$ may not work for high-dimensional data as demonstrated in \cite{ledoit2002some, chen2010two}. Besides, the estimator $V_3$ has unnecessary terms, which slows down the convergence considerably when dimension of PMU data is high \cite{chen2010two, chen2012tests}. In such situations, to reduce the drawbacks, trace criterion \cite{chen2010two} is more suitable to the tests problem. Instead of estimating the population covariance matrix directly, a well defined distance measure exploiting the difference among data flow $\left\{ {{{\bf{Z}}_1},{{\bf{Z}}_2}, \cdots ,{{\bf{Z}}_q}} \right\}$ is conducted, that is, the trace-based distance measure between ${{\bf{\Sigma }}_s}$ and ${{\bf{\Sigma }}_t}$ is
\begin{equation}
\label{eqB3}
{\mathop{\rm tr}\nolimits} \left\{ {{{\left( {{{\bf{\Sigma }}_s} - {{\bf{\Sigma }}_t}} \right)}^2}} \right\} = {\mathop{\rm tr}\nolimits} \left( {{\bf{\Sigma }}_s^2} \right) + {\mathop{\rm tr}\nolimits} \left( {{\bf{\Sigma }}_t^2} \right) - 2{\mathop{\rm tr}\nolimits} \left( {{{\bf{\Sigma }}_s}{{\bf{\Sigma }}_t}} \right),
\end{equation}
where ${\mathop{\rm tr}\nolimits} \left(  \cdot  \right)$ is the trace operator. Instead of estimating ${\mathop{\rm tr}\nolimits} \left( {{\bf{\Sigma }}_s^2} \right)$, ${\mathop{\rm tr}\nolimits} \left( {{\bf{\Sigma }}_t^2} \right)$,  and ${\mathop{\rm tr}\nolimits} \left( {{{\bf{\Sigma }}_s}{{\bf{\Sigma }}_t}} \right)$ by sample covariance matrix based estimators, we develop an unbiased estimator according to the U-statistics \cite{lee1990u} which allow a minimum-variance unbiased estimator to be derived from each unbiased estimator of an estimable parameter for large classes of probability distributions \cite{Cox1974Theoretical}. Specially, for $l = \left\{ {s,t} \right\} \in \Omega  = \left\{ {1 \le s,t \le q,s \ne t} \right\}$,
\begin{small}
\begin{eqnarray}
\label{eqB4}
{{{A}}_l} &=& \frac{1}{{{n_g}\left( {{n_g} - 1} \right)}}\sum\limits_{i \ne j} {{{\left( {{\bf{z}}_{li}^{'}{{\bf{z}}_{lj}}} \right)}^2}} \nonumber \\
&-& \frac{2}{{{n_g}\left( {{n_g} - 1} \right)\left( {{n_g} - 2} \right)}}\sum\limits_{i,j,k}^ *  {{\bf{z}}_{li}^{'}{{\bf{z}}_{lj}}{\bf{z}}_{lj}^{'}{{\bf{z}}_{lk}}}  \\
 &+& \frac{1}{{{n_g}\left( {{n_g} - 1} \right)\left( {{n_g} - 2} \right)\left( {{n_g} - 3} \right)}}\sum\limits_{i,j,k,h}^ *  {{\bf{z}}_{li}^{'}{{\bf{z}}_{lj}}{\bf{z}}_{lk}^{'}{{\bf{z}}_{lh}}} \nonumber
\end{eqnarray}
\end{small}is proposed to estimate ${\mathop{\rm tr}\nolimits} \left( {{\bf{\Sigma}}_l^2} \right)$. It is noted that $\sum\nolimits^{*}$ represents summation over mutually distinct indices. For example, $\sum\nolimits_{i,j,k}^ *$ says summation over the set $\left\{ {\left( {i,j,k} \right):i \ne j,j \ne k,k \ne i} \right\}$. Similarly, the estimator for ${\mathop{\rm tr}\nolimits} \left( {{{\bf{\Sigma }}_s}{{\bf{\Sigma }}_t}} \right)$ can be expressed as
\begin{small}
\begin{eqnarray}
\label{eqB5}
{C_{st}} &=& \frac{1}{{{n_g^2}}}\sum\limits_i {\sum\limits_j {{{\left( {{\bf{z}}_{si}^{'}{{\bf{z}}_{tj}}} \right)}^2}} } \nonumber \\
&-& \frac{1}{{\left( {{n_g} - 1} \right){n_g^2}}}\sum\limits_{i,h}^* {\sum\limits_j {{\bf{z}}_{si}^{'}{{\bf{z}}_{tj}}{\bf{z}}_{tj}^{'}{{\bf{z}}_{sh}}} } \nonumber \\
 &-& \frac{1}{{\left( {{n_g} - 1} \right){n_g^2}}}\sum\limits_{i,l}^* {\sum\limits_j {{\bf{z}}_{ti}^{'}{{\bf{z}}_{sj}}{\bf{z}}_{sj}^{'}{{\bf{z}}_{th}}} }   \\
 &+& \frac{1}{{\left( {{n_g} - 1} \right)^2{n_g^2}}}\sum\limits_{i,h}^* {\sum\limits_{j,k}^* {{\bf{z}}_{si}^{'}{{\bf{z}}_{tj}}{\bf{z}}_{sk}^{'}{{\bf{z}}_{th}}} } \nonumber .
\end{eqnarray}
\end{small}

The test statistic that measures the distance between ${{\bf{\Sigma }}_s}$ and ${{\bf{\Sigma }}_t}$ is
\begin{equation}
\label{eqB6}
{V_{st}} = {A_s} + {A_t} - {C_{st}}.
\end{equation}
Then the proposed test statistic can be expressed as:
\begin{equation}
\label{eqB7}
{V_1} = \frac{1}{{{q}\left( {{q} - 1} \right)}}\sum\limits_{\left\{ {s,t} \right\} \in \Omega } {{V_{st}}}.
\end{equation}

As $p,{n_g} \to \infty$, the asymptotic normality \cite{chen2012tests} of the test statistic \eqref{eqB6} is presented in the following:
\begin{theorem}
\label{thm1}
Let $\sigma _{st}^2 = \frac{1}{n_g}\left( {{A_s} + {A_t}} \right)$. Assuming the following conditions:
\begin{enumerate}
\item For any $k$ and $l \in \left\{ {s,t} \right\}$, ${\mathop{\rm tr}\nolimits} \left( {{{\bf{\Sigma }}_k}{{\bf{\Sigma }}_l}} \right) \to \infty$ and \[{\mathop{\rm tr}\nolimits} \left\{ {\left( {{{\bf{\Sigma }}_i}{{\bf{\Sigma }}_j}} \right)\left( {{{\bf{\Sigma }}_k}{{\bf{\Sigma }}_l}} \right)} \right\} = O\left\{ {{\mathop{\rm tr}\nolimits} \left( {{{\bf{\Sigma }}_i}{{\bf{\Sigma }}_j}} \right){\mathop{\rm tr}\nolimits} \left( {{{\bf{\Sigma }}_k}{{\bf{\Sigma }}_l}} \right)} \right\}.\]
\item For $i = 1,2, \cdots, n_g$, ${{\bf{z}}^{\left( i \right)}}$ are independent and identically distributed $p$-dimensional vectors with finite $8th$ moment.
\end{enumerate}
Under above conditions,
\[L = \frac{{{V_{st}}}}{{\sigma _{st}}}\mathop  \to \limits^d \mathcal{N}\left( {0,1} \right)\]
\end{theorem}

\begin{proposition}
\label{prop}
For any $q \ge 2$, as $p,{n_g} \to \infty$, the proposed test statistic $V_1$ satisfies
\begin{equation}
\label{eqB8}
{V_1}\mathop  \to \limits^d  \mathcal{N}\left( {\mu ,{\sigma ^2}} \right),
\end{equation}
where $\mu \approx 0, \sigma ^2 = \mathop \sum \nolimits^* \sigma _{st}^2$.
\end{proposition}
Let $R = \frac{{{V_1}}}{{\sigma _{{V_1}}}} $, the false alarm rate (FAR) for the proposed test statistic can be represented as
\begin{eqnarray}
\label{eqB9}
{P_{{FAR}}} &=& P\left( {R > \alpha |{H_0}} \right) \nonumber \\
 &=& \int_R^\infty  {\frac{1}{{\sqrt {2\pi } }}\exp \left( {\frac{{ - {t^2}}}{2}} \right)} dt \nonumber \\
 &=& Q\left( R \right),
\end{eqnarray}
where $Q\left( x \right) = \int_x^\infty  {{1 \mathord{\left/ {\vphantom {1 {\sqrt {2\pi } }}} \right. \kern-\nulldelimiterspace} {\sqrt {2\pi } }}\exp \left( {{{ - {t^2}} \mathord{\left/ {\vphantom {{ - {t^2}} 2}} \right. \kern-\nulldelimiterspace} 2}} \right)} dt$. For a desired FAR $\tau$, the associated threshold should be chosen such that \[\alpha  = {Q^{ - 1}}\left( \tau  \right) .\]
Otherwise, the detection rate (DR) can be denoted as
\begin{equation}
\label{DR}
{P_{{DR}}} = P \left( {R \ge {Q (\alpha) }|{H_1}} \right).
\end{equation}

It is noted that the computation complexity of proposed test statistic in \eqref{eqB8} is $O(\varepsilon n_g^4)$,  which limits its practical application. Here, we propose a effective approach to reducing complexity of the proposed test statistic from $O(\varepsilon n_g^4)$ to $O(\eta n_g^2)$ by principal component calculation and redundant computation elimination. For simplicity, the technical details are deferred to the Appendix B.

\subsection{Continuous Learning of the Power System}

Based on the proposed multiple high-dimensional test \eqref{eqB7} in Section \ref{ProsedTest}, we propose a method in the continuous manner to indicate the state evaluation. Details are shown in the following:

Let \[{T_{trn}} = \left[ {T_{trn}^{11}, \cdots ,T_{trn}^{1q}, \cdots ,T_{trn}^{n1}, \cdots ,T_{trn}^{nq}} \right]\] be the total training period. It is presumed that the power system is under normal operation during time period $T_{trn}$. For $i=1,2,\cdots,n$, the collected PMU data flow \[\left\{ {{{\bf{Z}}_{i1}},{{\bf{Z}}_{i2}}, \cdots ,{{\bf{Z}}_{iq}}} \right\}\] is employed for continuous learning of the power system parameters, namely, mean and variance of the proposed test statistic, detection threshold in \eqref{eqB9}, and then power system state. Specifically,

\subsubsection{Estimating the relative magnitude and duration of the system event}
\label{faultfind}

Using the proposed test statistic in \eqref{eqB7}, a system event can be identified with several samples of PMU data when the system event indicator satisfies
\begin{equation}
\label{eqTau}
\left| {{V_1} - \mu } \right| \ge \gamma ,
\end{equation}
where $\mu$, $\gamma = 3\sigma$ are the system-dependent parameters which can be learned from explanatory historical PMU data in the training procedure. The relative magnitude of a system event equals the test statistic $V_1$. Given that a system event occurs in sample period $T_{test}$, for $j=1,2,\cdots,c$, denotes the test data flow as $\left\{ {{{\bf{Z}}_{i1}},{{\bf{Z}}_{i2}}, \cdots ,{{\bf{Z}}_{iq}}} \right\}$, the duration of the event can be roughly estimated by
\begin{equation}
\label{dur}
{T_{dur}} = \sum\limits_{j = 1}^c {q*T*{\omega _j}},
\end{equation}
where \[{\omega _j} = \left\{ {\begin{array}{*{20}{c}}
1,  &{\left| {{V_1} - \mu } \right| \ge \gamma }\\
0,  &{\left| {{V_1} - \mu } \right| < \gamma }
\end{array}} \right. .\]

\subsubsection{Determination of the most sensitive PMU}

According to the data analysis in \ref{faultfind}, the voltage event addressed on a power system can be identified. Then, the determination of the most sensitive PMU is another important part to be investigated with respect to a system event.

The fact that every fault has its own effect on a power system \cite{bollen2000understanding} stimulates us to find the location of most sensitive PMU. According to the data analysis in Section \ref{faultfind}, we are able to determine the time when a system event occurs, say, $T_1$. Assume that the power system operates under normal condition during the time period of $T_1 -1$ and there are $p$ types of influential factors during a sampling time $T_1$. Denote ${{{\bf{Z}}^{\left( {{i}} \right)}}} = \left\{ {{{\bf{Z}}_{i1}},{{\bf{Z}}_{i2}}, \cdots ,{{\bf{Z}}_{iq}}} \right\}$, ${{{\bf{Z}}^{\left( {{j}} \right)}}} = \left\{ {{{\bf{Z}}_{j1}},{{\bf{Z}}_{j2}}, \cdots ,{{\bf{Z}}_{jq}}} \right\}$ and ${{{\bf{Z}}^{\left( {{k}} \right)}}} = \left\{ {{{\bf{Z}}_{k1}},{{\bf{Z}}_{k2}}, \cdots ,{{\bf{Z}}_{kq}}} \right\}$ as the PMU data flow collected during sample time $T_1-2$, $T_1-1$ and $T_1$. For $l=1,2,\cdots,p$, the measured data of each factor are formed as a row vector ${\bf{c}}^{(T)}_l$. In order to reveal the most sensitive PMU, we form a factor matrix by duplicating $\kappa$ times for each factor ${\bf{c}}^{(T_1)}_l$, say,
\begin{equation}
\label{eqB10}
{{\bf{C}}^{\left( {{T_1}} \right)}} = {\left[ {\begin{array}{*{20}{c}}
{{\bf{c}}_l^{\left( {{T_1}} \right)}}\\
 \vdots \\
{{\bf{c}}_l^{\left( {{T_1}} \right)}}
\end{array}} \right]_{\kappa \times {N}}},
\end{equation}
where the parameter $N=q*n_g$, $\kappa = r \log p$ and $r$ is the rank of ${{{\bf{Z}}^{\left( {{j}} \right)}}}$. For $l=1,2,\cdots , p$, we can construct two expansion matrices for parallel data analysis, formulated by
\begin{equation}
\label{eqB11}
{{\bf{A}}^{\left( l \right)}_{1}} = \left[ {\begin{array}{*{20}{c}}
{{{\bf{Z}}^{\left( {{i}} \right)}}}\\
{{{\bf{C}}^{\left( {{T_1}} \right)}}}
\end{array}} \right], {{\bf{A}}^{\left( l \right)}_{2}} = \left[ {\begin{array}{*{20}{c}}
{{{\bf{Z}}^{\left( {{j}} \right)}}}\\
{{{\bf{C}}^{\left( {{T_1}} \right)}}}
\end{array}} \right].
\end{equation}
Substitute data flows ${{\bf{A}}_{1l}}$ and ${{\bf{A}}_{2l}}$ into the test statistic in \eqref{eqB7}, the location of most sensitive PMU data (denoted as $loc$) during the sample time $T_1$ can be expressed as
\begin{equation}
\label{eqB12}
loc = {\rm{index}}\left( {\mathop {\max }\limits_{l = 1,2, \cdots ,p} \left( {V_1^{\left( l \right)}} \right)} \right),
\end{equation}
where ${\rm{index}}\left( {{x_{j}}} \right) = j$.

For the readers' convenience, the technological process of the proposed test statistic for power state evaluation are summarized in the following:
\begin{table}[!ht]
\begin{tabularx}{0.47\textwidth}{l}
\toprule
\bf{Implementation of the proposed state evaluation indicator.}
\\
\midrule
1): Off-line training period (System-dependent parameters learning): \\
\quad 1a): collect the PMU data and represent them using \eqref{eq1}; \\
\quad 1b): calculate the test statistic of the data flow using \eqref{eqB7}; \\
\quad 1c): calculate mean and variance of the proposed test statistic; \\
\quad 1d): determine the event indicator threshold $\gamma$ using \eqref{eqTau}; \\
2): Online power state indicating: \\
\quad 2a): acquire the test data flow: ${{\bf{Z}}_{j1}}, \cdots ,{{\bf{Z}}_{jq}}, j=1,2,\cdots,c$;  \\
\quad 2b): calculate the test statistic of the data flow using \eqref{eqB7}; \\
\quad 2c): determine whether there is an event using \eqref{eqTau}; \\
\qquad \qquad if no event detected: \\
\qquad \qquad \qquad  add the test data flow into history data;   \\
\qquad \qquad \qquad  go back to the step 1a); \\
\qquad \qquad else:  \\
\qquad \qquad \qquad go to step 2d); \\
\quad 2d): Determine the relative magnitude, duration and location of the \\
\quad \ system event using \eqref{eqB7}, \eqref{dur} and \eqref{eqB12}, respectively; \\
3): Performance evaluation: \\
\quad 3a): FAR \eqref{eqB9} and DR \eqref{DR} analysis; \\
\quad 3b): the effect of measurement noise analysis; \\
\quad 3c): the effect of parameter $q$ analysis. \\
\bottomrule
\end{tabularx}
\end{table}

So far, the power state evaluation by the proposed test statistic is established. Case studies to evaluate the practical performance of the proposed test statistic will be depicted in detail in the following section.

\section{Case Studies and Discussions}
\label{sec:case}
The proposed test statistic for power state evaluation are numerically evaluated by the power network benchmarks, namely the IEEE 30-, 118-bus system, a Polish 2383-bus system \cite{kavasseri2011joint}, and a real 34-PMU system. For the synthetic data, the admittance matrices and the underlying power system states are generated by MATPOWER package \cite{zimmerman2011matpower}. It is noted that the measurement noise is simulated as uncorrelated Gaussian or Gama distribution with a standard deviation per component of 0.05 for voltages \cite{bollen2000understanding, zimmerman2011matpower}. We report results from case studies which are designed to evaluate the performance of the proposed test for the power state evaluation in the following.

\subsection{Effect of measurement noise on the power state evaluation}
\label{noise}

As discussed in Section \ref{Sec:1}, PMUs offer highly accurate measurements (i.e., voltage, current phasors and frequency) when operating under steady-state conditions. Significant error creeps into the measurements, however, while operating under transient conditions. The amount of measurement error varies from one manufacturer to another due to the difference in the method used to calculate the output quantities. This poses a serious question to the power state indicator's ability to monitor dynamics of a power system at the time of disturbances when we have no prior knowledge about measurement noise. So the effect of measurement noise on the power state evaluation is firstly studied as follows.

Assuming that the power system operates under normal state, we first investigate the effect of measurement noise and window size on the state evaluation using synthetic data. With respect to the proposed test in \eqref{eqB7}, we generate $p$-dimensional data independent multivariate data models using the linearized measurement model in \eqref{eqC3}. Let ${{\bf{z}}_0}$ be the initial state of the power system. The nominal significance level \cite{bollen2000understanding} of the data and parameter $q$ are set to $5\%$ and 10, respectively. For $i=1,2, \cdots , n_g$, we consider two scenarios regarding the innovation random vector ${{\bf{z}}^{\left(i \right)}}$:

\begin{enumerate}
\item ${{\bf{z}}^{\left(i \right)}}$ are $p$-dimensional normal random vector with mean ${{\bf{z}}_0}$ and variance ${\rm{diag}}\left\{ {0.05{\bf{z}}_0} \right\}$.
\item ${\bf{z}}^{\left( i \right)} = {\left[ {z_{1}^{\left( i \right)}, \cdots ,z_{p}^{\left( i \right)}} \right]^{'}}$ consist of independent random variables ${z_{j}^{\left( i \right)}}$ which are standard Gamma$({\bf{z}}_0,0.2236)$ + 0.7764${\bf{z}}_0$ random variables.
\end{enumerate}

It is noted that the proposed test statistic imposes no restriction on the relationship between the data dimension and sample size. To mimic the buses deployed in the power system, we have $p \in \left\{ {30, 118, 2383} \right\}$. A wide range of sample window sizes are denoted as $n_g \in \left\{ {30, 100, 300, 1000, 2500} \right\}$. The simulation results (Tab. \ref{table:powers1} and Tab. \ref{table:powers2}) reported in this section are based on 1000 independent Monte Carlo simulations.

\begin{table}[!ht]
\caption{{\label{table:powers1}} DR and FAR of the test statistics with GSN.}
\begin{tabularx}{0.47\textwidth}{lccllll p{0.6cm}p{0.6cm}p{0.6cm}p{0.6cm}p{0.6cm}p{0.6cm}p{0.6cm}}
\toprule
\multirow{1}{*}{}
& \multicolumn{2}{c}{$LR \ test$}  & \multicolumn{2}{c}{$CLR \ test$}  & \multicolumn{2}{c}{$Proposed \ test$} \\
 \cmidrule(lr){2-3}  \cmidrule(lr){4-5}  \cmidrule(lr){6-7}    
\multirow{1}{*}{$(p,n_g,q)$}
& \multicolumn{1}{c}{DR} & \multicolumn{1}{c}{FAR}
& \multicolumn{1}{c}{DR} & \multicolumn{1}{c}{FAR}
& \multicolumn{1}{c}{DR} & \multicolumn{1}{c}{FAR}
\\
\midrule
(30,30,10)     & 0.595 & 0.059 & 0.651 & 0.067 & 0.694 & 0.061  \\
(30,100,10)    & 0.742 & 0.064 & 0.899 & 0.061 & 0.912 & 0.058  \\
(30,300,10)    & 0.901 & 0.089 & 0.955 & 0.057 & 0.979 & 0.047  \\
(30,1000,10)   & 0.958 & 0.134 & 0.997 & 0.054 & 0.999 & 0.039  \\
(30,2500,10)   & 1     & 0.296 & 1     & 0.049 & 1     & 0.049  \\
(118,30,10)    & -     & -     & 0.924 & 0.047 & 0.985 & 0.059  \\
(118,100,10)   & -     & -     & 0.957 & 0.051 & 0.993 & 0.055  \\
(118,300,10)   & 0.995 & 0.149 & 0.993 & 0.053 & 1     & 0.049  \\
(118,1000,10)  & 1     & 0.390 & 1     & 0.048 & 1     & 0.045  \\
(118,2500,10)  & 1     & 0.483 & 1     & 0.045 & 1     & 0.043  \\
(2383,30,10)   & -     & -     & 0.991 & 0.063 & 0.995 & 0.058  \\
(2383,100,10)  & -     & -     & 1     & 0.055 & 1     & 0.053 \\
(2383,300,10)  & -     & -     & 1     & 0.051 & 1     & 0.050 \\
(2383,1000,10) & -     & -     & 1     & 0.046 & 1     & 0.048 \\
(2383,2500,10) & 1     & 0.891 & 1     & 0.047 & 1     & 0.049 \\
\bottomrule
\end{tabularx}
\end{table}

\begin{table}[!ht]
\caption{{\label{table:powers2}} DR and FAR of the test statistics with GMN.}
\begin{tabularx}{0.47\textwidth}{lccllll p{0.6cm}p{0.6cm}p{0.6cm}p{0.6cm}p{0.6cm}p{0.6cm}p{0.6cm}}
\toprule
\multirow{1}{*}{}
& \multicolumn{2}{c}{$LR \ test$}  & \multicolumn{2}{c}{$CLR \ test$}  & \multicolumn{2}{c}{$Proposed \ test$} \\
 \cmidrule(lr){2-3}  \cmidrule(lr){4-5}  \cmidrule(lr){6-7}    
\multirow{1}{*}{$(p,n_g,q)$}
& \multicolumn{1}{c}{DR} & \multicolumn{1}{c}{FAR}
& \multicolumn{1}{c}{DR} & \multicolumn{1}{c}{FAR}
& \multicolumn{1}{c}{DR} & \multicolumn{1}{c}{FAR}
\\
\midrule
(30,30,10)     & 0.471 & 0.067 & 0.553 & 0.073 & 0.476 & 0.069  \\
(30,100,10)    & 0.660 & 0.163 & 0.643 & 0.075 & 0.775 & 0.067  \\
(30,300,10)    & 0.791 & 0.289 & 0.816 & 0.067 & 0.891 & 0.066  \\
(30,1000,10)   & 0.958 & 0.334 & 0.894 & 0.060 & 0.953 & 0.063  \\
(30,2500,10)   & 0.996 & 0.596 & 0.934 & 0.057 & 0.989 & 0.055  \\
(118,30,10)    & -     & -     & 0.801 & 0.066 & 0.885 & 0.059  \\
(118,100,10)   & -     & -     & 0.879 & 0.059 & 0.967 & 0.063  \\
(118,300,10)   & 0.932 & 0.349 & 0.942 & 0.063 & 0.995 & 0.056  \\
(118,1000,10)  & 0.999 & 0.875 & 0.970 & 0.056 & 1     & 0.052  \\
(118,2500,10)  & 1     & 0.977 & 0.998 & 0.051 & 1     & 0.055  \\
(2383,30,10)   & -     & -     & 0.947 & 0.062 & 0.984 & 0.061  \\
(2383,100,10)  & -     & -     & 0.983 & 0.058 & 0.999 & 0.060  \\
(2383,300,10)  & -     & -     & 1     & 0.059 & 1     & 0.054  \\
(2383,1000,10) & -     & -     & 1     & 0.049 & 1     & 0.052  \\
(2383,2500,10) & 1     & 1     & 1     & 0.046 & 1     & 0.048  \\
\bottomrule
\end{tabularx}
\end{table}

The simulation results in Tab. \ref{table:powers1} and Tab. \ref{table:powers2} show that DR of the covered test statistics increase as the dimension and sample sizes become larger. Many entries of the DR of the tests approach 1 in both of the scenarios of Gaussion distributed noise (GSN) and Gama distributed noise (GMN). Besides, as shown in the Table I and Table II, FAR of the proposed test converge to the nominal 5\% quite rapidly with $p$ and $n_g$ increase for both GSN and GMN; Meanwhile, the convergence of the FAR to the nominal level for GSN is slower than GMN. On the other hand, the LR test is not applicable for $p \ge n_g$ and CLRT test shows slower convergence than the proposed test. In other words, the proposed test statistic has more accurate DR and robust FAR in a quite wide range of dimensionality and distributions while the LR test and the CLR test are vulnerable to variation in the data dimension and noise distribution. This could be understood as the proposed test is both asymptotic and nonparametric.

\subsection{Effect of the parameter $q$ on the power state evaluation}
\label{q_effect}
\begin{figure}[htbp]
\centering
\subfloat[]{ \label{fig_q1}
\includegraphics[width=0.5\columnwidth]{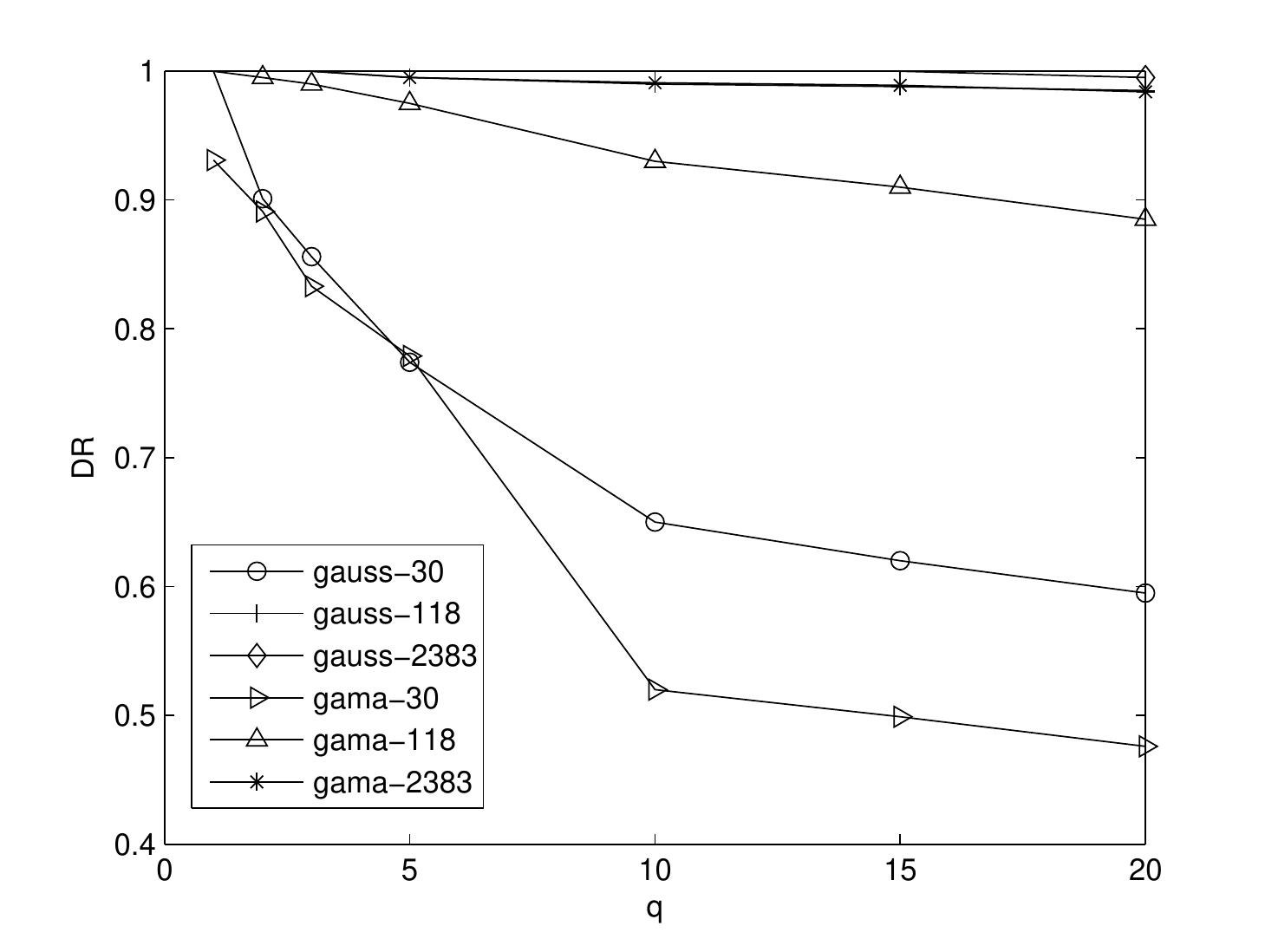}
}
\subfloat[]{ \label{fig_q2}
\includegraphics[width=0.5\columnwidth]{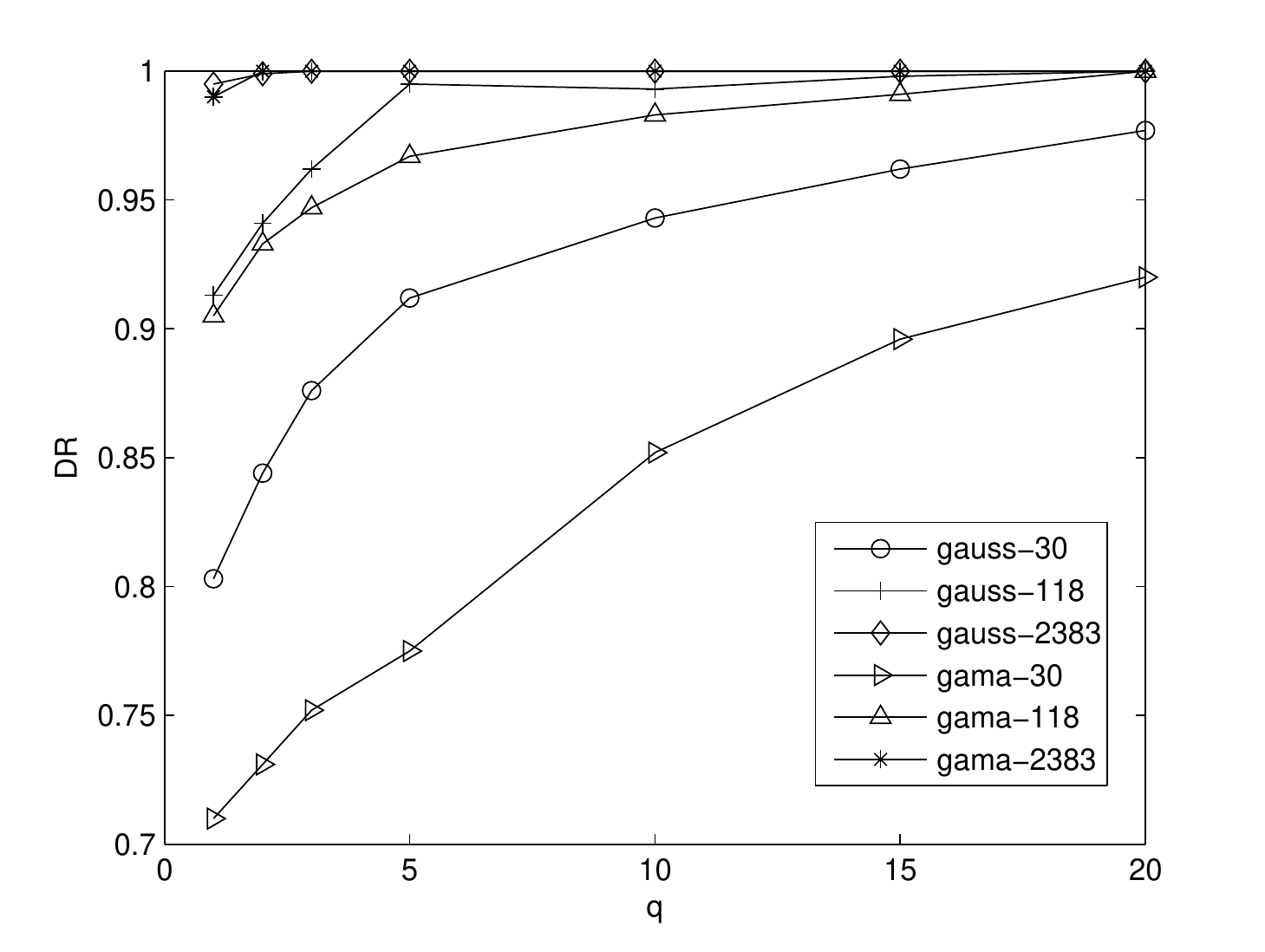}
}
\caption{Effect of the parameter $q$ on the power state evaluation.}
\label{fig_q}
\end{figure}

As depicted in the Section \ref{Sec:2}, the parameter $q$ is an important factor for state evaluation. More details illustrated by experimental data are shown in the following. We fix the total data size as 600, that is $q*n_g = 600$ in first experiment while setting the window size $n_g$ as 100 in the second one. Two kinds of measurement noise shown in the Section \ref{noise} are considered. It is noted that the notations $gauss-30$ and $gama-30$ in Fig. \ref{fig_q1} and Fig. \ref{fig_q2} mean that the measurement noise adopted are GSN and GMN with the number of PMUs at $p=30$, respectively. Similar definitions also work for other notations, i.e.,  $gauss-118$, $gama-118$, $gauss-2383$, and $gama-2383$.

Fig. \ref{fig_q1} shows that the DR decreases as $q$ increases for the first experiment while Fig. \ref{fig_q2} illustrates that the DR shows a positive response to the increase of $q$ in the second one. The selection of a medium size $q$ is the trade-off between the DR and the realtime performance. In the rest of the experiments, we set the parameter $q$ as $q=5$.

\subsection{Online Power State Evaluation Using the Synthetic Data}
\label{OPS}

The performance of the power state evaluation using the proposed test statistic is evaluated by the simulated data generated from IEEE 30-bus, IEEE 118-bus, and a polish 2383-bus system, respectively. The specific details of the systems are referred to the case30.m, case118.m and case2383.m in the Matpower package and Matpower 5.1-User's Manual \cite{zimmerman2015matpower}. In the simulations, changes on the active load of each bus are considered as potential factors. Besides, each change of a factor is described as a signal. Three kinds of signals that affect the operating state of the test system are considered. For simplicity, the signals for each factor are shown in Tab. \ref{table:case1}, \ref{table:case2} and \ref{table:case3}. $\rho$ denotes the number of P-V nodes in the test systems and is chosen on a random basis. For the sake of simplicity, the case studies based on the IEEE 118-bus system are presented below. The results generated from IEEE 30-bus system and the Polish 2383-bus system are deferred to the Appendix C.

\begin{table}[!ht]
\begin{center}
\caption{{\label{table:case1}}Signal Type I: Voltage Dip}
\begin{tabular}{ccc}
\toprule
Bus & Duration & Active Load (MW)\\
\midrule
\multirow{3}{*}{$\rho $}   & \multicolumn{1}{c}{$t=1\sim 300$} & \multicolumn{1}{c}{40.0} \\ & \multicolumn{1}{c}{$t=301\sim 600$} & \multicolumn{1}{c}{80} \\ & \multicolumn{1}{c}{$t=601\sim 1000$} & \multicolumn{1}{c}{120} \\
\hline
 Others & $t=1\sim 1000$ & Unchanged \\
\bottomrule
\end{tabular}
\end{center}

\end{table}

\begin{table}[!ht]

\begin{center}
\caption{{\label{table:case2}}Signal Type II: Voltage Swell}
\begin{tabular}{ccc}
\toprule
Bus & Duration & Active Load (MW)\\
\midrule
\multirow{5}{*}{$\rho $}   & \multicolumn{1}{c}{$t=1\sim 300$} & \multicolumn{1}{c}{-10.0} \\ & \multicolumn{1}{c}{$t=301\sim 540$} & \multicolumn{1}{c}{-25.1} \\ & \multicolumn{1}{c}{$t=541\sim 780$} & \multicolumn{1}{c}{-39.3} \\
& \multicolumn{1}{c}{$t=781\sim 900$} & \multicolumn{1}{c}{-62.7} \\ & \multicolumn{1}{c}{$t=901\sim 1000$} & \multicolumn{1}{c}{-75.3} \\
\hline
 Others & $t=1\sim 1000$ & Unchanged \\
\bottomrule
\end{tabular}
\end{center}

\end{table}

\begin{table}[!ht]

\begin{center}
\caption{{\label{table:case3}}Signal Type III: Voltage Dip and Swell}
\begin{tabular}{ccc}
\toprule
Bus & Duration & Active Load (MW)\\
\midrule
\multirow{4}{*}{$\rho $}   & \multicolumn{1}{c}{$t=1\sim 300$} & \multicolumn{1}{c}{10.0} \\ & \multicolumn{1}{c}{$t=301\sim 600$} & \multicolumn{1}{c}{60.0} \\ & \multicolumn{1}{c}{$t=601\sim 900$} & \multicolumn{1}{c}{120.0} \\
& \multicolumn{1}{c}{$t=901\sim 1000$} & \multicolumn{1}{c}{35.0} \\
\hline
 Others & $t=1\sim 1000$ & Unchanged \\
\bottomrule
\end{tabular}
\end{center}

\end{table}

\begin{figure}[htbp]
\centering
\subfloat[Parameter learning with GSN.]{ \label{fig118gama}
\includegraphics[width=0.5\columnwidth]{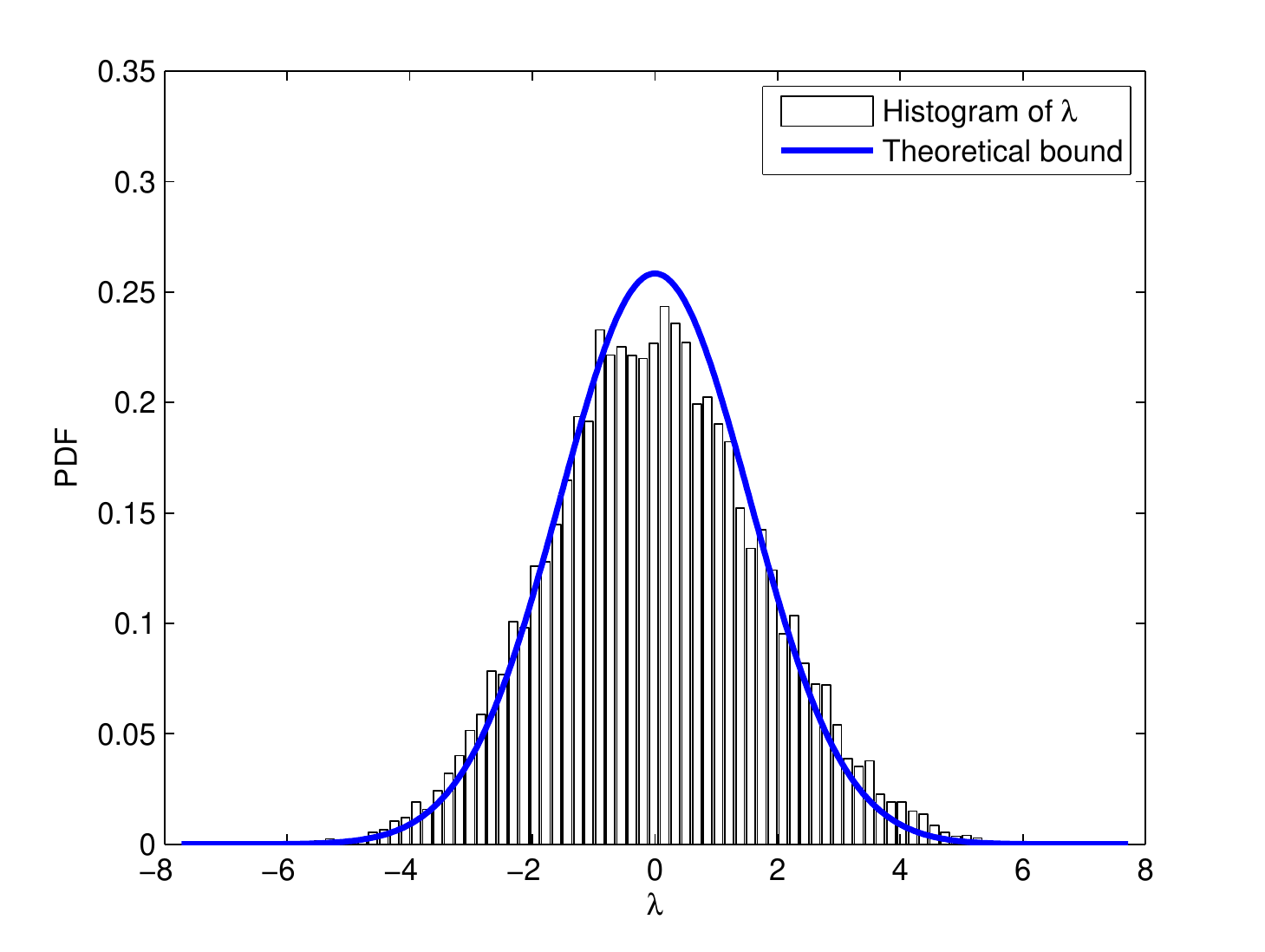}
}
\subfloat[Parameter learning with GMN.]{ \label{fig118gauss}
\includegraphics[width=0.5\columnwidth]{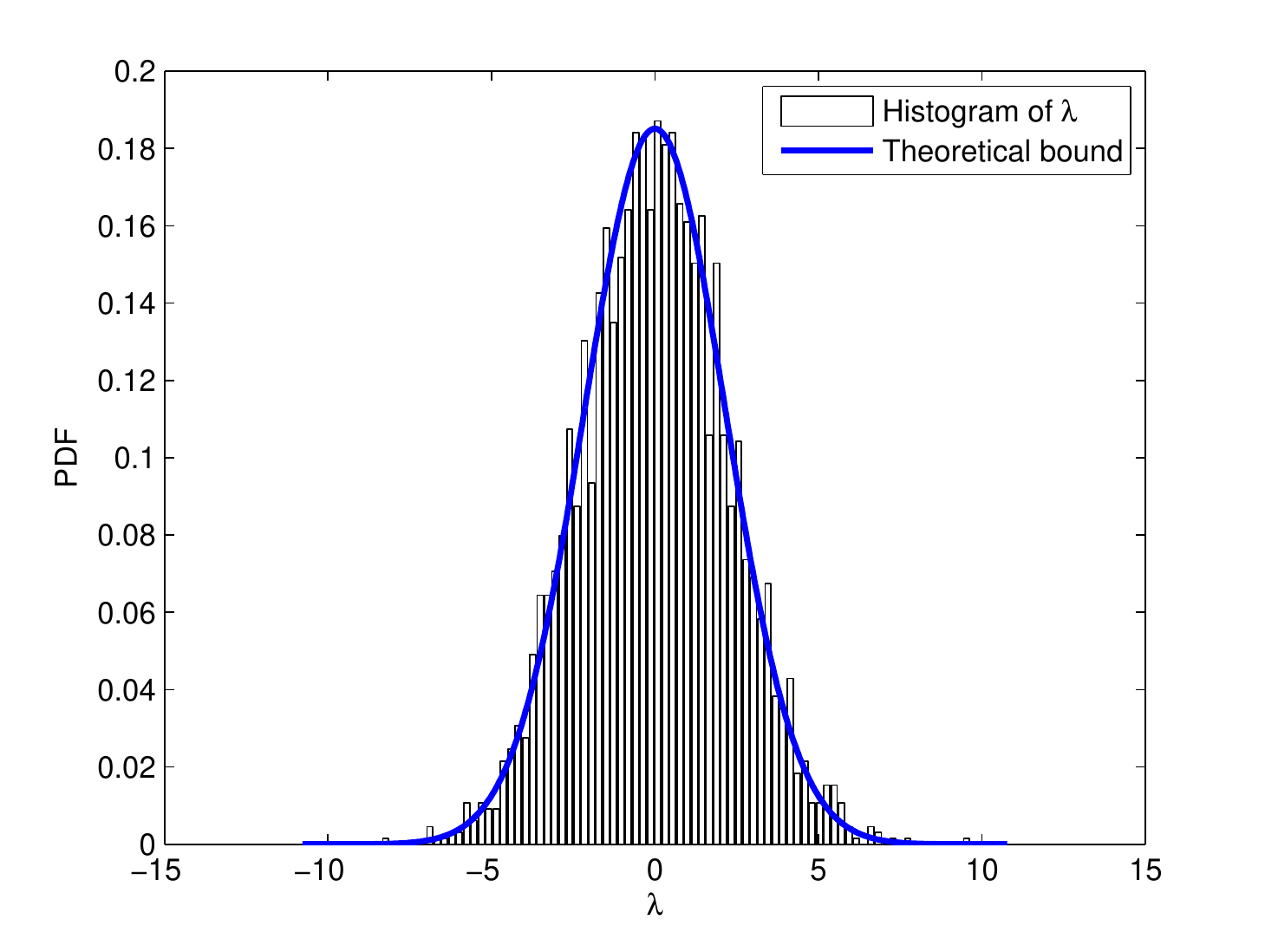}
}
\caption{Parameter learning of IEEE 118-bus system}
\label{fig118learn}
\end{figure}

The signals are generated in load of P-V node $\rho = 63$ for the case of IEEE 118-bus system. During the training period, 5 minutes of data are collected when the system is under normal condition. Let $p = 118, n_g = 100, q = 5$. Two kinds of measurement noise shown in the Section \ref{noise} are considered. As shown in the Section \ref{ProsedTest}, the proposed test statistic satisfies $\lambda \mathop  \to \limits^d \mathcal{N}\left( {0,1} \right)$. The theoretical bound in Fig.\ref{fig118learn} is the probability density function (PDF) of $\lambda$. Fig.\ref{fig118learn} shows that the mean and variance of $\lambda$ fit fabulously with theoretical ones.

\begin{figure}[htbp]
\subfloat[]{\label{fig118gs1}
\includegraphics[width=0.24\textwidth]{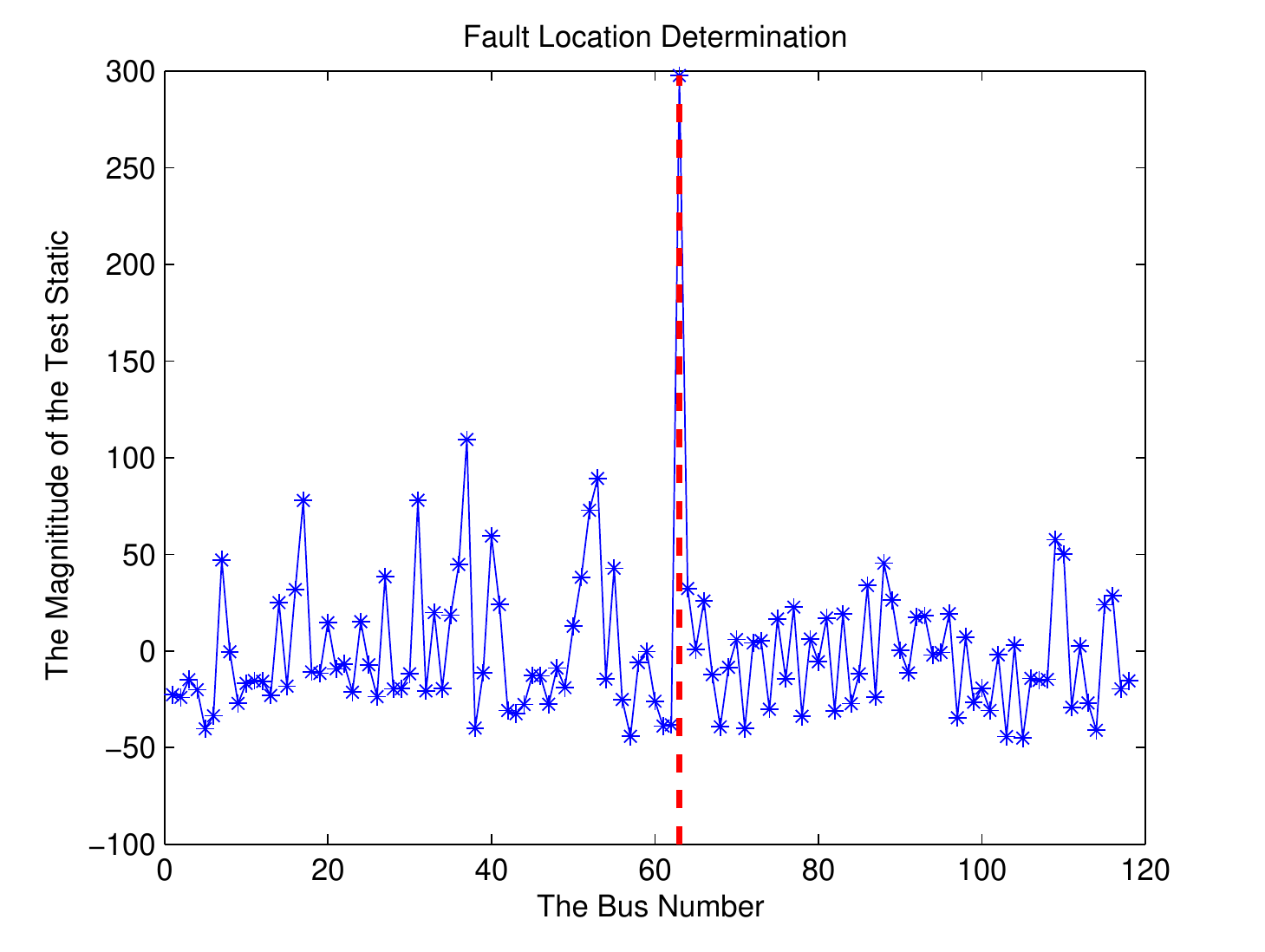}
}
\subfloat[]{\label{fig118gm1}
\includegraphics[width=0.5\columnwidth]{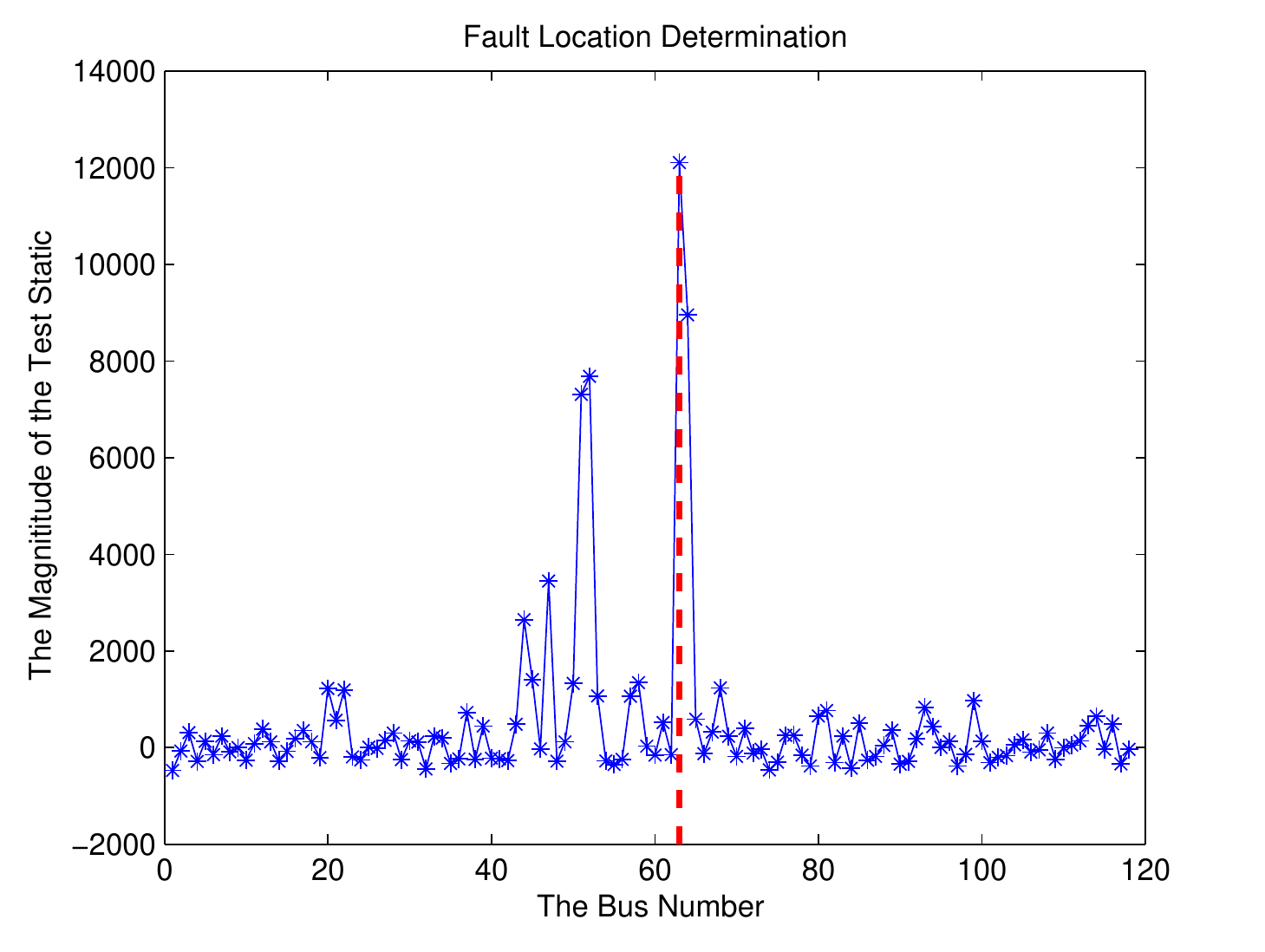}
}\\
\subfloat[]{\label{fig118gs2}
\includegraphics[width=0.5\columnwidth]{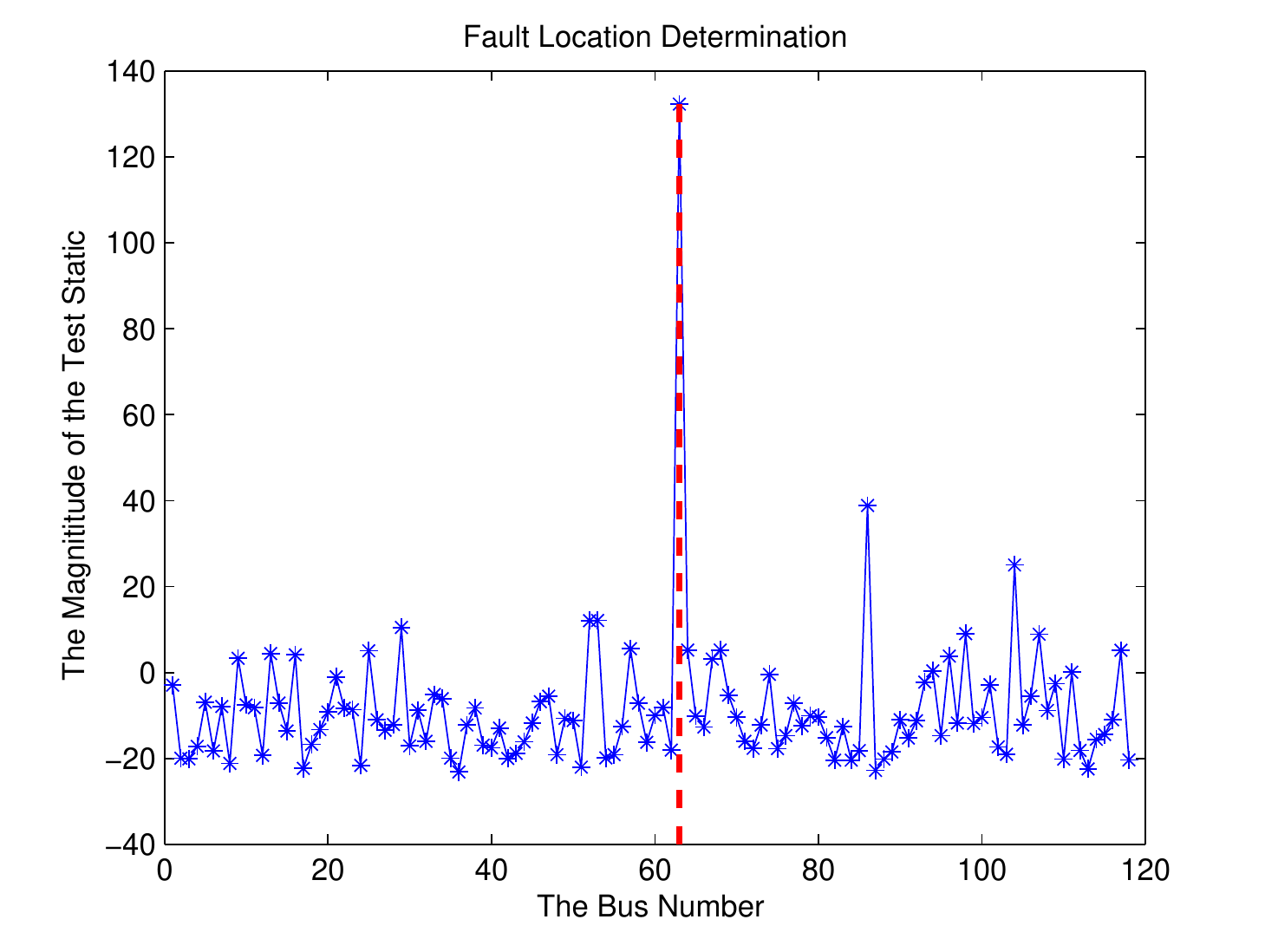}
}
\subfloat[]{\label{fig118gm2}
\includegraphics[width=0.5\columnwidth]{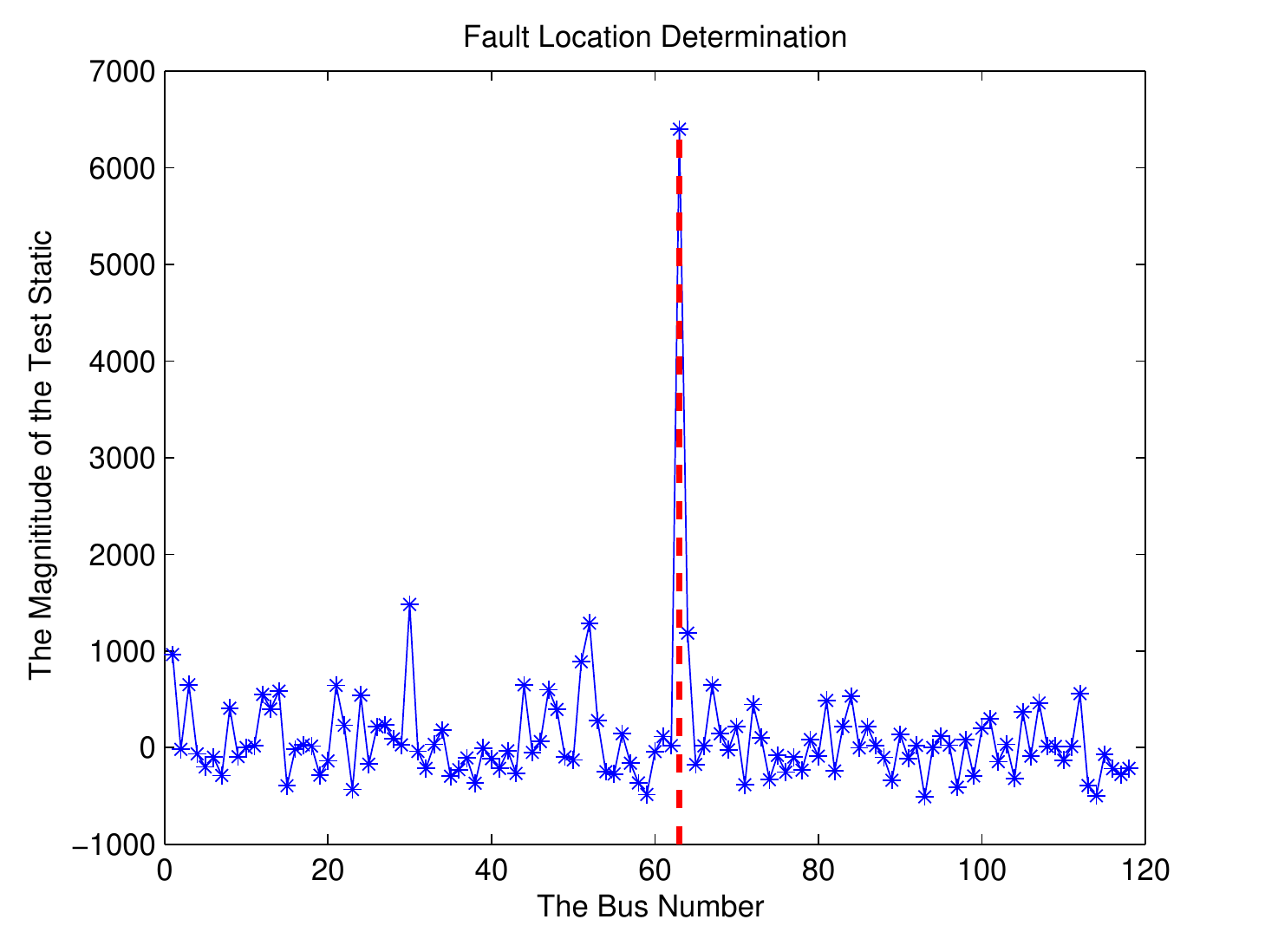}
}\\
\subfloat[]{\label{fig118gs3}
\includegraphics[width=0.5\columnwidth]{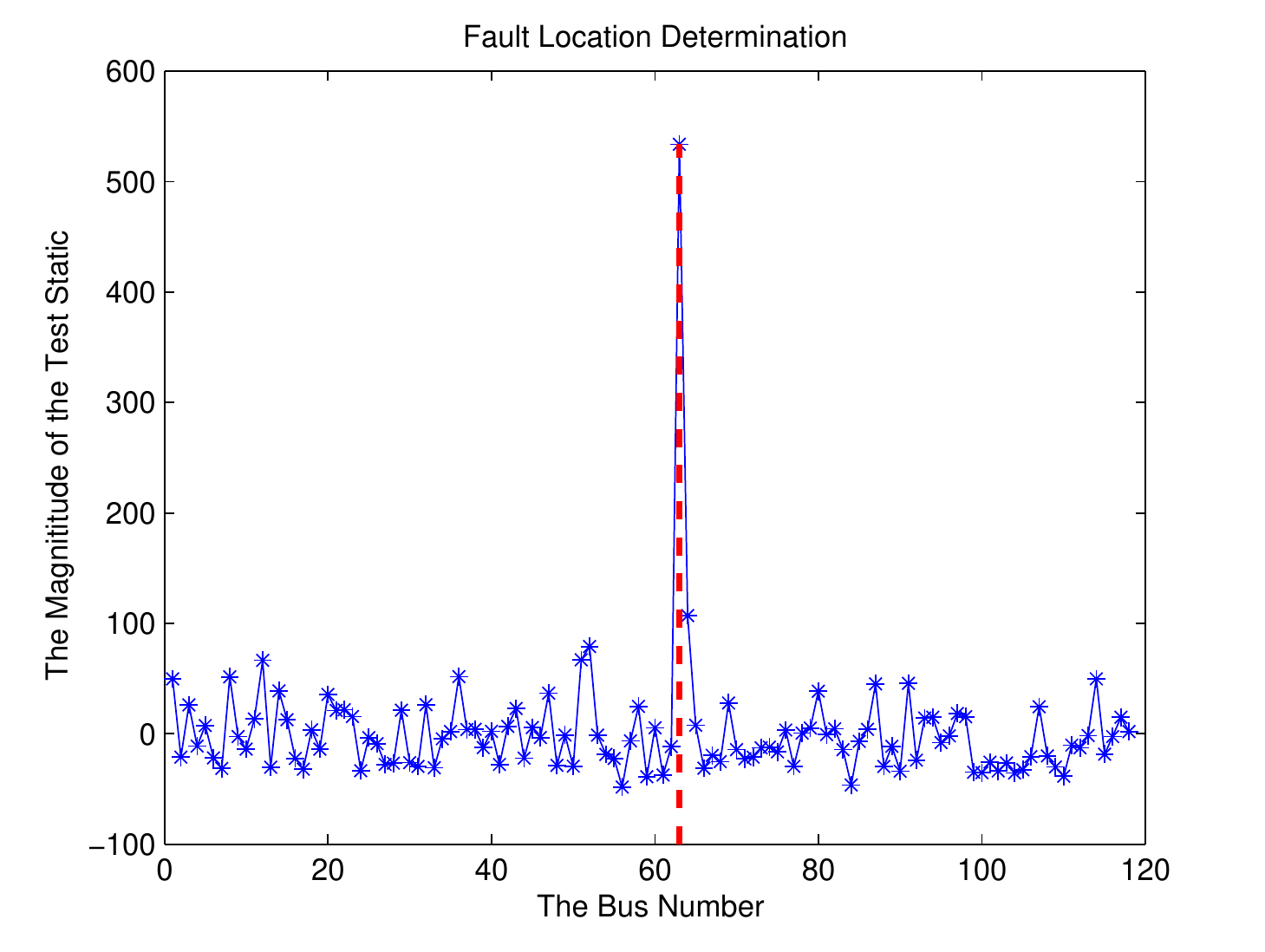}
}
\subfloat[]{\label{fig118gm3}
\includegraphics[width=0.5\columnwidth]{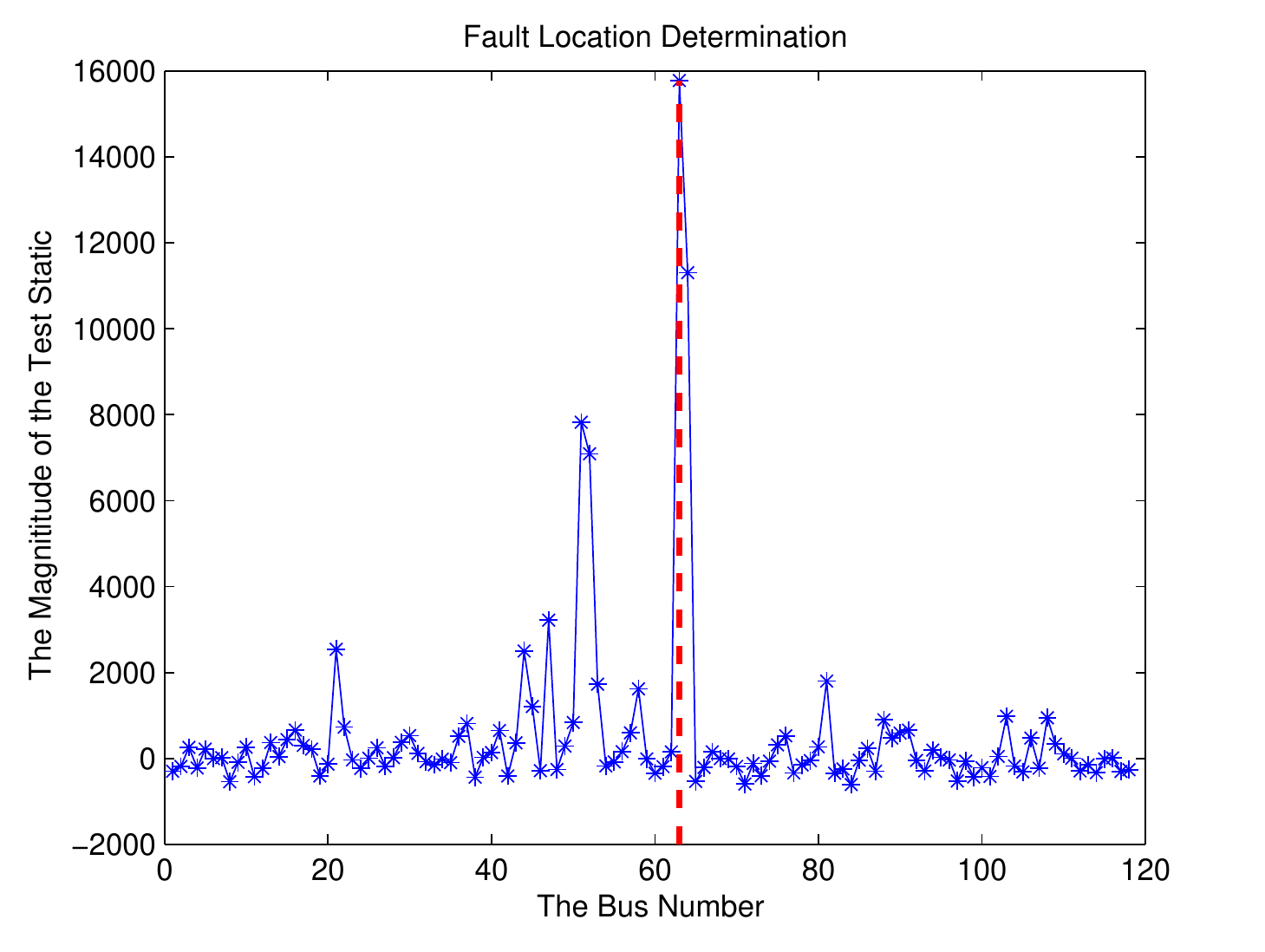}
}
\caption{Determination of most sensitive bus for IEEE 118-bus system. (The system are effected by Type I signal with GSN, Type I signal with GMN, Type II signal with GSN, Type II signal with GNN, Type III signal with GSN and Type III signal with GMN, respectively.)}
\label{figcase118}
\end{figure}

The power state evaluation begins at 301th $s$.  60 seconds of data are collected. Three kinds of system events are generated in load of 63$th$ bus from 320s to 340s, respectively. According to the results in Fig.\ref{fig118learn} and event indicators \eqref{eqTau} and \eqref{dur}, we can know that the event occurs at 301s and the actual duration of the signals can be calculated as $t_dur = 1000/(q*n_g)*10 = 20s$. Based on the above analysis, we can then determine the location of the most sensitive bus using \eqref{eqB12}. The results in Fig.\ref{figcase118} demonstrate that the 63$th$ bus is the most sensitive bus in presence of all three kinds system events when the measurement noise is set as GSN or GMN.

\subsection{A Real Data Analysis}

In this section, we evaluate the efficacy of the proposed test statistic for the power system state. For the experiments shown in the following, the real power flow is of a chain-reaction fault that happened in the China power grids in 2013. The PMU number, the sample rate, and the total sample time are $p = 34$, $K = 50 Hz$ and $284s$, respectively. The chain-reaction fault happened from $t=65.4s$ to $t=73.3s$.

\begin{figure}[htbp]
{
\includegraphics[width=0.475\textwidth]{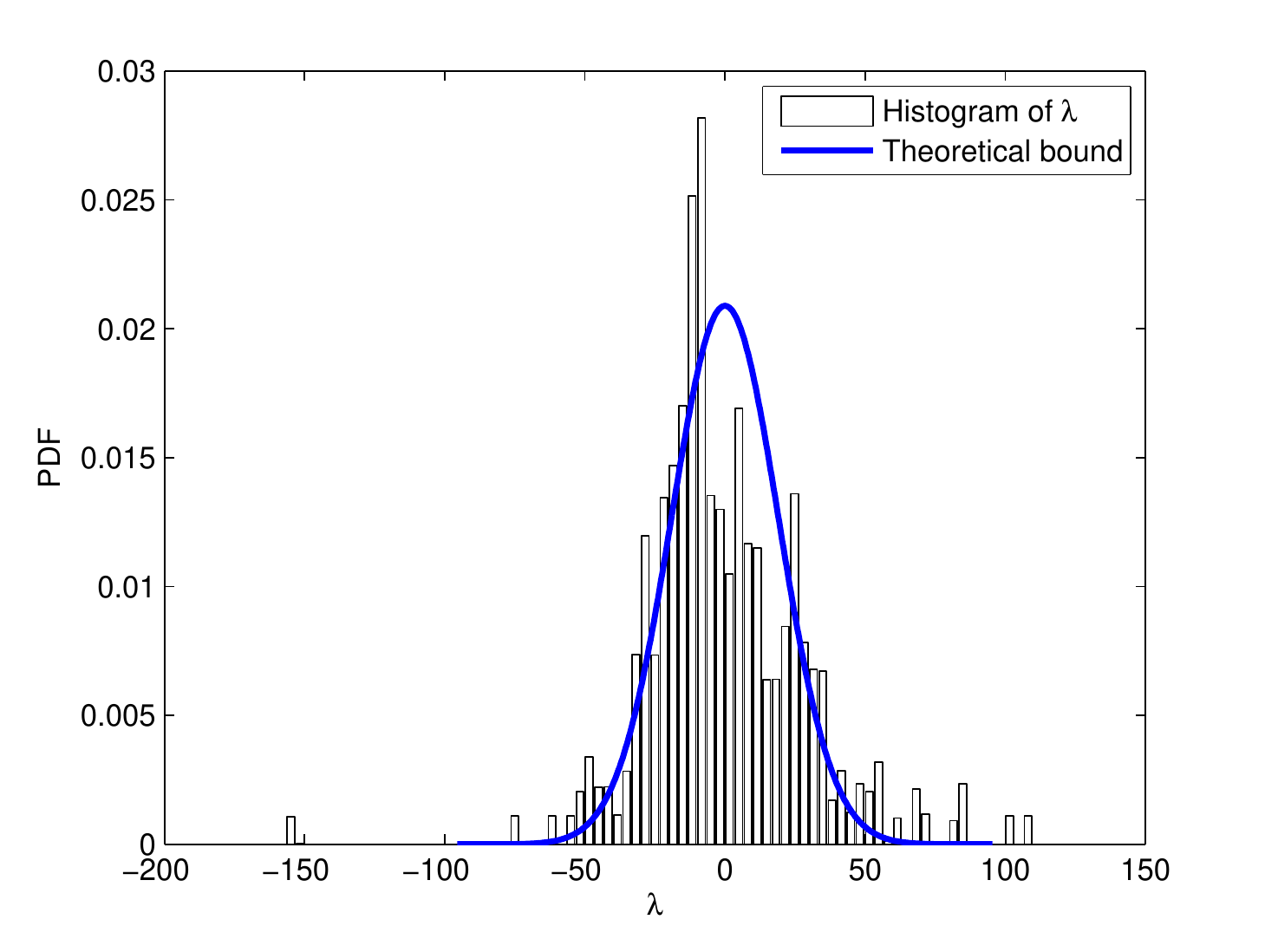}
}
\caption{{\label{real_data_learning}} Parameter learning of the real 34-PMU system.}
\end{figure}

Let $q = 5, n_g = 50$. Fig.\ref{real_data_learning} shows that the mean and variance of $\lambda$ agree well with theoretical ones. Based on the results in Fig.\ref{real_data_learning} and event indicators \eqref{eqTau} and \eqref{dur}, the occurrence time and the actual duration of the event can be identified as $t_0 = 65s$ and $t_dur \approx 8s$, respectively. Similar to the data analysis above, we can then determine the location of the most sensitive bus using \eqref{eqB12}. The results illustrate that 17$th$ and 18$th$ PMU are the most sensitive PMUs, which are in accordance with the actual accident situation.



\begin{figure}[htbp]
{
\includegraphics[width=0.475\textwidth]{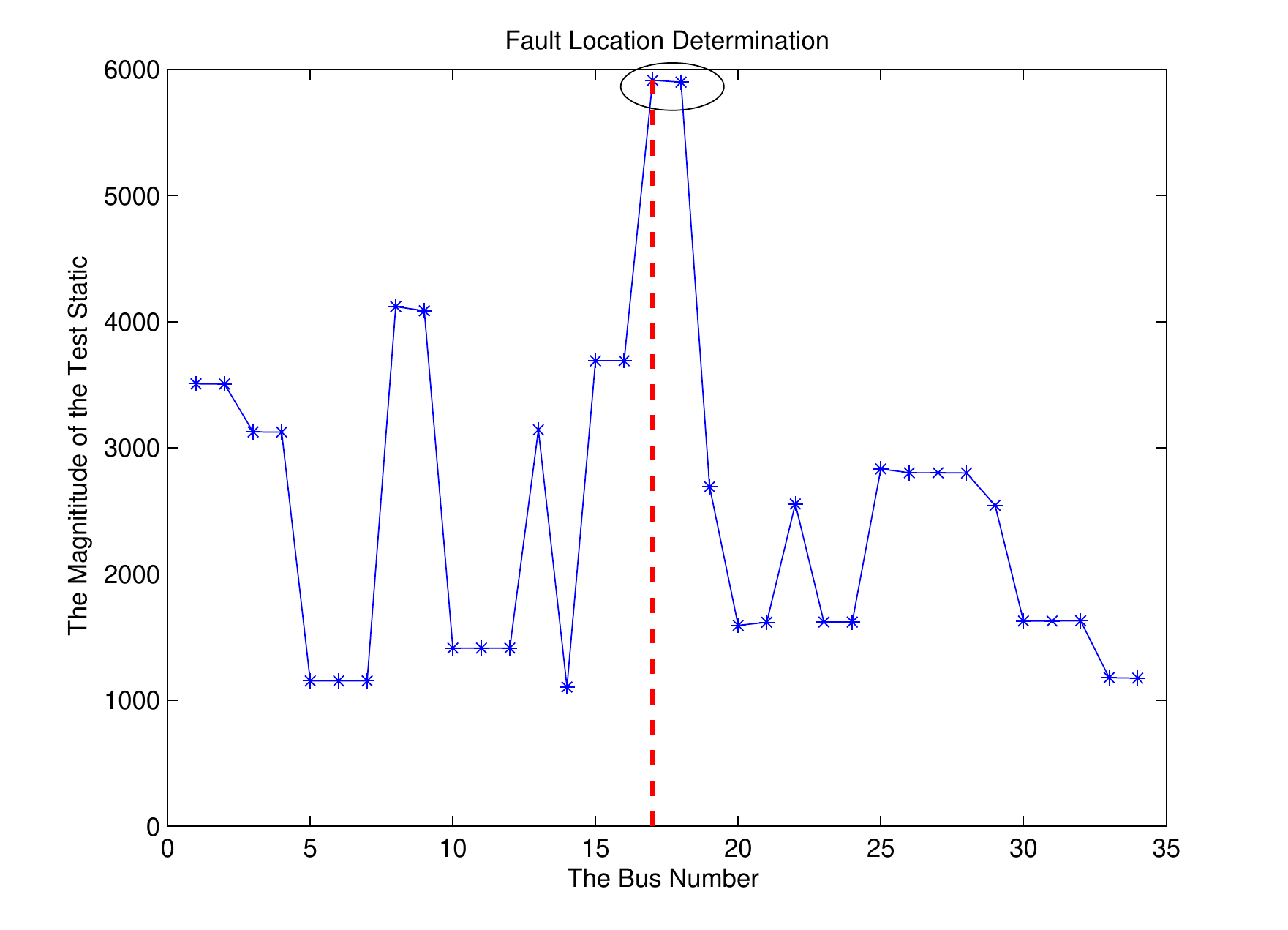}
}
\caption{{\label{real_data_loc}} Determination of most sensitive bus for the real 34-PMU system.}
\end{figure}

\subsection{More Discussions in Details}

The general objective of the wide deployment of the PMU installation equipments is to eventually make a transition from the conventional supervisory control and data acquisition based measurement system to a more superior measurement system that will use synchronized measurements collected from geographically distant locations and increase the power state evaluation by monitoring a wide area of the power system in real time. The main problem raised from large power systems is how to tackle the tough situation when massive large dimensional measurements get together. This paper enhances the connection between big data analysis and large scale power system, by providing a comprehensive analysis for massive streaming PMU data. In the case studies above, we demonstrate the efficacy of the proposed power state indicator, including the parameter selection and power state indication, under different scenarios. However, there are still some interesting details left.

Results in Section \ref{noise} show that our proposed non-parameter power state indicator work well in two types of measurement noise (GSN and GMN); More types of noise are also suggested and studied. Besides, as shown in Section \ref{Sec:2} and Section \ref{q_effect}, the parameter $q$ acts as an important role in power state evaluation. There should be some connections between phenomena in power systems and $q$ in mathematics. Due to the limit space, we leave that for our future work. Moreover, three kinds of signals that affect the operating state of the test system are considered in Section \ref{OPS}. In a real power system, those signals are mainly caused by the three types of system events: control input changes, initial condition changes or system topology changes \cite{bollen2000understanding}; Those signals would occur irregularly in the modern power system where enormous new energy resources and renewable energy sources exist. The results in Section \ref{OPS} demonstrate that our advanced power state indicator could reveal the relative magnitude, duration and location of any system event. More detailed case studies need to be developed to harvest the maximum benefit of the proposed power state indicator for monitoring, protection, and control of a power system under all operating conditions (i.e, the scenario where there exist more types of signals or more complicated signals).

As mentioned in Section \ref{Sec:1}, the new metric proposed in this paper is based on multiple high dimensional covariance matrix tests. The advantage of the asymptotic property of the proposed test statistic is studied and employed. The results in Fig. \ref{fig118learn} and Fig. \ref{real_data_learning} show that the p.d.f of the proposed test statistic fits fabulously within the theoretical bound for synthetic data while the performance is a little bit worse for real data. Therefore, deducing a more precise bound for the proposed test statistic is necessary in future work.

This work develops an essential mathematical framework with a sound theoretical guarantee. Besides, the favorable properties (i,e., parameter-independence, flexibility and low computation complexity) of the proposed method would benefit not only readers with electric backgrounds but a large number of interested readers (i.e., from computer science, bioengineering, or applied mathematics) who encounter similar problems.


\section{Conclusion}
\label{sec:Conclusion}

Motivated by the immediate demands of big data analysis for large scale smart grid, this paper proposes a real time data-driven method to indicate the state evaluation from massive streaming PMU data. First, we represent the PMU data as a sequence of large random matrices. This is a crucial part for the power state evaluation as it turnes the big PMU data into tiny data for practical use. Rather than employing the raw PMU data or window-truncated data, a comprehensive analysis of PMU data flow, namely, the multiple high-dimensional test, is then proposed to indicate the power system state. The case studies based on synthetic data and real data show that the proposed test statistic can accurately reveal the relative magnitude, duration, and location of a system event in polynomial time.

The current work provides a fundamental exploration of data analysis for massive streaming PMU data. Much more attention is to be paid along this research direction, such as classification of power events from massive streaming PMU data. It is also noted that this work is a data-driven method which is a new substitute for power system state estimation. The combination of power system scenario analysis and the data driven methods is encouraged to be further investigated for a better understanding of the power system state.

\section{Acknowledgement}
\label{sec:Acknowledgement}

The authors would like to thank the anonymous reviewers for a critical reading of the manuscript, which led to a significant improvement of this paper.


\begin{appendices}

\section{The traditional high-dimensional covariance matrix tests}
\label{append1}

For the readers' convenience, we give a brief description of the likelihood ratio (LR) test statistic and correction of the likelihood ratio (CLR) test statistic in the following.

Let $\sum\limits_{g = 1}^q {{n_g}}  = n$ be the total sample size, ${{{\bf{\bar z}}}_g} = \sum\limits_{k = 1}^{{n_g}} {{{\bf{z}}_{gk}}},$
\begin{eqnarray}
\label{AP:eq1}
{{\bf{Y}}_g} &=& \frac{1}{{{n_g - 1}}} \sum\limits_{k = 1}^{{n_g}} {\left( {{{\bf{z}}_{gk}} - {{{\bf{\bar z}}}_g}} \right){{\left( {{{\bf{z}}_{gk}} - {{{\bf{\bar z}}}_g}} \right)}^{'}}}  \\
{\bf{Y}} &=& \sum\limits_{g = 1}^q {{{\bf{Y}}_g}}. \nonumber
\end{eqnarray}
The LR test \cite{bai1996effect} for testing hypothesis equation (5) is
\begin{equation}
\label{eqB100}
{V_2} = \frac{{\prod\limits_{g = 1}^q {{{\left| {{{\bf{Y}}_g}} \right|}^{\frac{1}{2}{N_g}}}} }}{{{{\left| {\bf{Y}} \right|}^{\frac{1}{2}M}}}},
\end{equation}
where
\[{N_g} = {n_g} - 1, \quad M = {N_1} + {N_2} +  \cdots  + {N_q} = n - q.\]

It is noted that the calculation of the numerator and denominator of $V_2$ will lead to overflow as $n_g$ becomes large. To overcome the overflow difficulty, a CLR test \cite{bai2009corrections} for testing the equality of more population covariance matrices is shown as follows. Let
\begin{equation}
\label{eqB101}
{V_{2h}} = \frac{{{{\left| {{{\bf{Y}}_1} + {{\bf{Y}}_2} +  \cdots  + {{\bf{Y}}_{h - 1}}} \right|}^{\frac{1}{2}\left( {{N_1} + {N_2} +  \cdots  + {N_{h - 1}}} \right)}}{{\left| {{{\bf{Y}}_h}} \right|}^{\frac{1}{2}{N_h}}}}}{{{{\left| {\bf{Y}} \right|}^{\frac{1}{2}M}}}},
\end{equation}
where $h=2,3, \cdots, q$. Then ${V_2} = \prod\limits_{h = 2}^q {{V_{2h}}}$.

The CLR test statistic is
\begin{equation}
\label{eqB102}
{V_3} = \sum\limits_{h = 2}^q { - \frac{2}{{{N_1} + {N_2} +  \cdots  + {N_{h - 1}}}}} \log {T_{1h}} - pf\left( {{y_{1h}},{y_{2h}}} \right),
\end{equation}
where
\[{y_{1h}} = \frac{p}{{{N_1} + {N_2} +  \cdots  + {N_{h - 1}}}},{y_{2h}} = \frac{p}{{{N_{h}}}}\]
and
\begin{small}
\begin{eqnarray}
f\left( {{y_{1}},{y_{2}}} \right) &=& \frac{{{y_{1}} + {y_{2}} - {y_{1}}{y_{2}}}}{{{y_{1}}{y_{2}}}}\log \left( {\frac{{{y_{1}} + {y_{2}}}}{{{y_{1}} + {y_{2}} - {y_{1}}{y_{2}}}}} \right) \nonumber \\
 &+& \frac{{y_1^2\left( {1 - {y_2}} \right)\log \left( {1 - {y_2}} \right) + y_2^2\left( {1 - {y_1}} \right)\log \left( {1 - {y_1}} \right)}}{{{y_1}{y_2}\left( {{y_1} + {y_2}} \right)}} \nonumber \\
 &-& \frac{{{y_{1}}}}{{{y_{1}} + {y_{2}}}}\log \frac{{{y_{1}}}}{{{y_{1}} + {y_{2}}}} - \frac{{{y_{2}}}}{{{y_{1}} + {y_{2}}}}\log \frac{{{y_{2}}}}{{{y_{1}} + {y_{2}}}}. \nonumber
\end{eqnarray}
\end{small}

%


\section{Computation Aspect of the proposed test statistic}
\label{append2}

From Equation (7)-(9), we know that the computational complexity of calculating the test statistics ${A_s}$, ${A_t}$ and ${C_{st}}$ are $O(\varepsilon _1 n_g^4)$, $O(\varepsilon _2 n_g^4)$ and $O(\varepsilon _3 n_g^4)$, respectively. With the increasing scale of PMU deployment and the increasing complexity of issues addressed by it, which is a newly raised challenge for the power state evaluation and quality control of a power system. Here, we propose a lower complexity method to calculate ${A_s}$, ${A_t}$, and ${C_{st}}$ by by redundant computation elimination and principal component calculation. Technical details are elaborated in the following.

\subsection{Redundant Computation Elimination}

We first consider eliminating the index-wise redundant computation during calculating the term ${A_{l,\left\{ {l = s,t} \right\}}}$.

Let
\[{A_{l1}} = \frac{1}{{{n_g}\left( {{n_g} - 1} \right)}}\sum\limits_{i \ne j} {{{\left( {{\bf{z}}_{li}^{'}{{\bf{z}}_{lj}}} \right)}^2}}, \]
\[{A_{l2}} = \frac{2}{{{n_g}\left( {{n_g} - 1} \right)\left( {{n_g} - 2} \right)}}\sum\limits_{i,j,k}^ *  {{\bf{z}}_{li}^{'}{{\bf{z}}_{lj}}{\bf{z}}_{lj}^{'}{{\bf{z}}_{lk}}},  \]
and
\[{A_{l3}} = \frac{1}{{{n_g}\left( {{n_g} - 1} \right)\left( {{n_g} - 2} \right)\left( {{n_g} - 3} \right)}}\sum\limits_{i,j,k,h}^ *  {{\bf{z}}_{li}^{'}{{\bf{z}}_{lj}}{\bf{z}}_{lk}^{'}{{\bf{z}}_{lh}}}. \]
It is easy to find that indices $i,j,k,l$ in $A_{l1}$, $A_{l2}$ and $A_{l3}$ are invariant with respect to the swapping places.
Let
\[\begin{array}{l}
{\Omega _1} = \left\{ {\left\{ {i,j} \right\}:1 \le i,j \le n_g,i \ne j} \right\},\\
{\Omega _2} = \left\{ {\left\{ {i,j,k} \right\}:1 \le i,j,k \le n_g,i \ne j \ne k} \right\},\\
{\Omega _3} = \left\{ {\left\{ {i,j,k,h} \right\}:1 \le i,j,k,h \le n_g,i \ne j \ne k \ne h} \right\}.
\end{array}\]
Specially, we are to determine unrepeated sets of the indices from $\Omega _1$, $\Omega _2$ and $\Omega _3$ when calculating $A_{l1}$, $A_{l2}$ and $A_{l3}$. Following the permutations and combinations principle in \cite{hall1983permutations}, the unrepeated ensembles can be expressed as
\[\begin{array}{l}
{{\dot \Omega }_1} = \left\{ {\left\{ {i,j} \right\}:2 \le i \le n_g,1 \le j \le i - 1} \right\},\\
{{\dot \Omega }_2} = \left\{ {\left\{ {i,j,k} \right\}:3 \le i \le n_g,2 \le j \le i - 1,1 \le k \le j - 1} \right\},\\
{{\dot \Omega }_3} = \left\{ \begin{array}{l}
\left\{ {i,j,k,h} \right\}:\left\{ {4 \le i \le n_g,3 \le j \le i - 1} \right\} \cup \\
\qquad \qquad \quad \left\{ {2 \le k \le j - 1,1 \le h \le k - 1} \right\}
\end{array} \right\}.
\end{array}\]

Let $Q_{n_g}^r = {{{n_g}!} \mathord{\left/
 {\vphantom {{{n_g}!} {\left( {n_g - r} \right)!}}} \right.
 \kern-\nulldelimiterspace} {\left( {n_g - r} \right)!}}$.
Then, $A_{l1}$, $A_{l2}$ and $A_{l3}$ can be expressed by

\[{A_{l1}} = \frac{2}{{Q_{n_g}^2}}\sum\limits_{\{ i,j\}  \in {{\dot \Omega }_1}} {{{\left( {{\bf{z}}_{li}^{'}{{\bf{z}}_{lj}}} \right)}^2}} ,\]
\[{A_{l2}} = \frac{6}{{Q_{n_g}^3}}\sum\limits_{\{ i,j,k\}  \in {{\dot \Omega }_2}} {{\bf{z}}_{li}^{'}{{\bf{z}}_{lj}}{\bf{z}}_{lj}^{'}{{\bf{z}}_{lk}}}, \] and
\[{A_{l3}} = \frac{{24}}{{Q_{n_g}^4}}\sum\limits_{\{ i,j,k,h\}  \in {{\dot \Omega }_3}} {{\bf{z}}_{li}^{'}{{\bf{z}}_{lj}}{\bf{z}}_{lk}^{'}{{\bf{z}}_{lh}}} .\]

As a result, the computational complexity of calculating $A_{l1}$, $A_{l2}$, and $A_{l3}$ is reduced by a factor of  $1/2$ , $1/6$, and $1/24$ compared with direct manipulation, respectively.

Besides, we notice that manipulation of  $A_{l1}$, $A_{l2}$, and $A_{l3}$ is completed in sequence and this manipulation is inefficient because of the repeated vector multiplication operations. For instance, vector multiplication ${{\bf{z}}_{li}^{'}{{\bf{z}}_{lj}}}$ is repeated many times when calculating $A_{l1}$, $A_{l2}$ and $A_{l3}$. This kind of repeated calculation can be avoided by the following steps.

Let ${{\bf{Z}}^l}$  be voltage-relevant matrix whose elements are \[Z_{ij}^l = {\bf{z}}_{li}^{'}{{\bf{z}}_{lj}}, {\{ i,j\}  \in {{\dot \Omega }_1}}.\]
Then $A_{l1}$, $A_{l2}$ and $A_{l3}$ can be equivalently denoted as
\[{A_{l1}} = \frac{2}{{Q_{n_g}^2}}\sum\limits_{\{ i,j\}  \in {{\dot \Omega }_1}} {{{\left( {Z_{ij}^l} \right)}^2}}, \]
\[{A_{l2}} = \frac{6}{{Q_{n_g}^3}}\sum\limits_{\{ i,j,k\}  \in {{\dot \Omega }_2}} {Z_{ij}^lZ_{jk}^l} ,\] and
\[{A_{l3}} = \frac{{24}}{{Q_{n_g}^4}}\sum\limits_{\{ i,j,k,h\}  \in {{\dot \Omega }_3}} {Z_{ij}^lZ_{kh}^l} .\]

The aforementioned equivalent expression means that we can compute ${{\bf{z}}_{li}^{'}{{\bf{z}}_{lj}}}$ only once during the progress in calculating $A_{l1}$, $A_{l2}$ and $A_{l3}$. Thus the computing time is reduced to $1/ n_g^2 $ of the conventional calculation of $A_{l2}$ and $A_{l3}$.

Similarly, the computation burden of calculating ${C_{s,t}}$ can be also alleviated by repeating the steps above. Here we only provide the result.  ${C_{s,t}}$ can be equivalently denoted as
\begin{eqnarray}
{C_{s,t}} &=& \frac{2}{{{n_g^2}}}\sum\limits_{\{ i,j\}  \in {{\dot \Omega }_1}} {{{\left( {{Y_{ij}}} \right)}^2}} \nonumber \\
 &-& \frac{{12}}{{n_gQ_{n_g}^2}}\sum\limits_{\{ i,j,k\}  \in {{\dot \Omega }_2}} {\left( {{Y_{ij}}Y_{jk}^{'} + Y_{ij}^{'}{Y_{jk}}} \right)} \nonumber \\
 &+& \frac{{24}}{{{{\left( {Q_{n_g}^2} \right)}^2}}}\sum\limits_{\{ i,j,k,h\}  \in {{\dot \Omega }_3}} {{Y_{ij}}{Y_{kh}}}, \nonumber
\end{eqnarray}
where $Y_{ij} = {\bf{z}}_{si}^{'}{{\bf{z}}_{tj}}, {\{ i,j\}  \in {{\dot \Omega }_1}}.$

\subsection{Principal Component Calculation}

Let $A = B + C$, where $A,B,C$ are positive random variables. Let $n_g$ be a large positive number, say, 100. If the condition that $C/B < 1/n_g$ is satisfied, then $B$ is called the principal component of $A$. Then we introduce the principal component calculation.

It can be noted that the magnitude of voltage measurements are positive, that is,
\[{{{Z}^l_{kj}}} > 0 , \left\{ {i,j} \right\} \in \left\{ {{{\dot \Omega }_1} \cup {{\dot \Omega }_2} \cup {{\dot \Omega }_3}} \right\}, \]
then
\[\sum\limits_{\{ i,j\}  \in {{\dot \Omega }_1}} {{{\left( {Z_{ij}^l} \right)}^2}}  > \sum\limits_{\{ i,j,k\}  \in {{\dot \Omega }_2}} {Z_{ij}^lZ_{jk}^l}  > \sum\limits_{\{ i,j,k,h\}  \in {{\dot \Omega }_3}} {Z_{ij}^lZ_{kh}^l}. \]
For $n_g\gg1$, divide ${A_{l2}}$ and ${A_{l3}}$  by ${A_{l1}}$, respectively, we can get
\begin{equation}
\label{eqD1}
\begin{array}{l}
\frac{{{A_{l2}}}}{{{A_{l1}}}} = \frac{{\frac{6}{{Q_{n_g}^3}}\sum\limits_{\{ i,j,k\}  \in {{\dot \Omega }_2}} {Z_{ij}^lZ_{jk}^l} }}{{\frac{2}{{Q_{n_g}^2}}\sum\limits_{\{ i,j\}  \in {{\dot \Omega }_1}} {{{\left( {Z_{ij}^l} \right)}^2}} }} < \frac{3}{{n_g - 2}} \ll 1,\\
\frac{{{A_{l3}}}}{{{A_{l1}}}} = \frac{{\frac{{24}}{{Q_{n_g}^4}}\sum\limits_{\{ i,j,k,h\}  \in {{\dot \Omega }_3}} {Z_{ij}^lZ_{kh}^l} }}{{\frac{2}{{Q_{n_g}^2}}\sum\limits_{\{ i,j\}  \in {{\dot \Omega }_1}} {{{\left( {Z_{ij}^l} \right)}^2}} }} > \frac{{12}}{{\left( {n_g - 2} \right)\left( {n_g - 3} \right)}} \ll 1.
\end{array}
\end{equation}
From \eqref{eqD1}, it is known that ${A_{l1}}$ is the principal component to be computed when computing ${A_{l}}$. We can get similar results when calculating ${C_{s,t}}$. Above all, the simplified test statistic can be represented as
\begin{equation}
\label{eqD2}
V_1 = \sum\limits_{\{ i,j\}  \in {{\dot \Omega }_1}} {\frac{2}{{Q_{n_g}^2}}\left( {{{\left( {Z_{ij}^1} \right)}^2} + {{\left( {Z_{ij}^p} \right)}^2}} \right) - \frac{2}{{{n_g^2}}}{{\left( {{Y_{ij}}} \right)}^2}}.
\end{equation}

Let $\varepsilon = \varepsilon_1 + \varepsilon_2 + \varepsilon_3$. It is noted that this kind of approximate computation will reduce the computation from $O(\varepsilon n_g^4)$ to $O(\eta n_g^2)$. The price paid for such an operation is that the simplified statistic in \eqref{eqD2} is no longer unbiased.

\section{Additional Case Study Results}
\label{re30_2383}

\begin{figure}[htbp]
\centering
\subfloat[Parameter learning with GSN.]{ \label{fig30gauss}
\includegraphics[width=0.5\columnwidth]{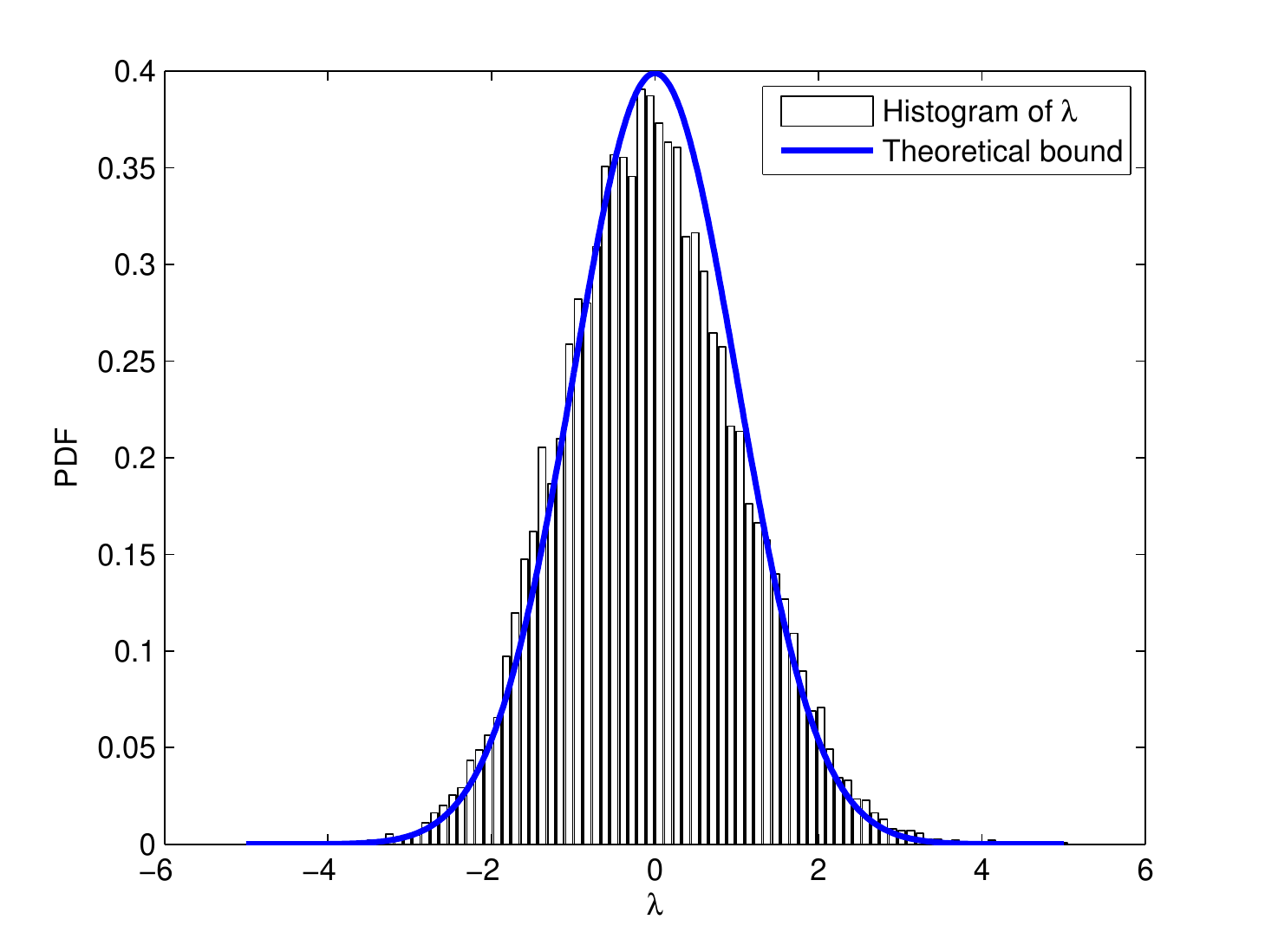}
}
\subfloat[Parameter learning with GMN.]{ \label{fig30gama}
\includegraphics[width=0.5\columnwidth]{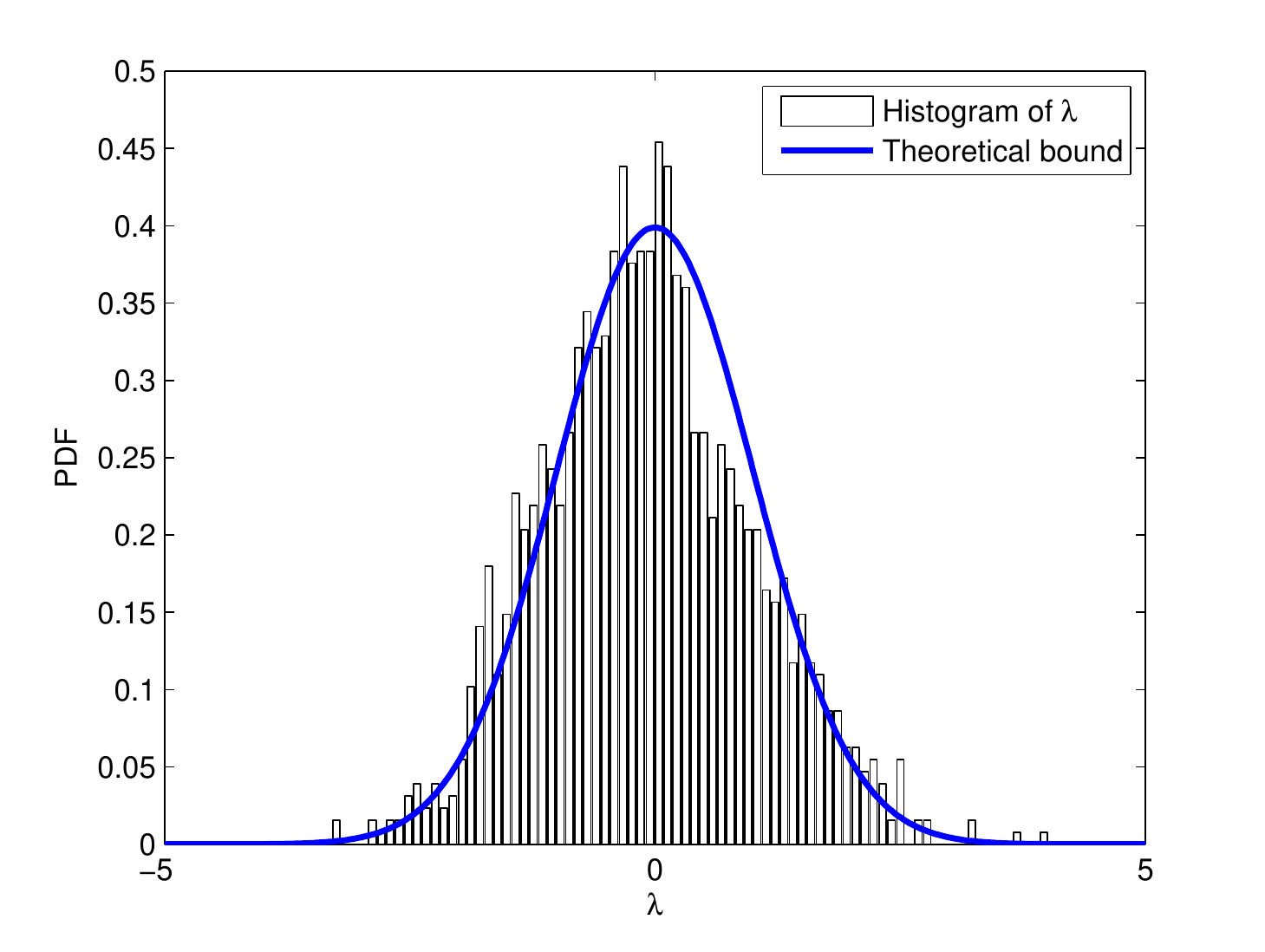}
}
\caption{Parameter learning of IEEE 30-bus system}
\label{fig30learn}
\end{figure}

\begin{figure}[htbp]
\subfloat[]{\label{fig30gs1}
\includegraphics[width=0.5\columnwidth]{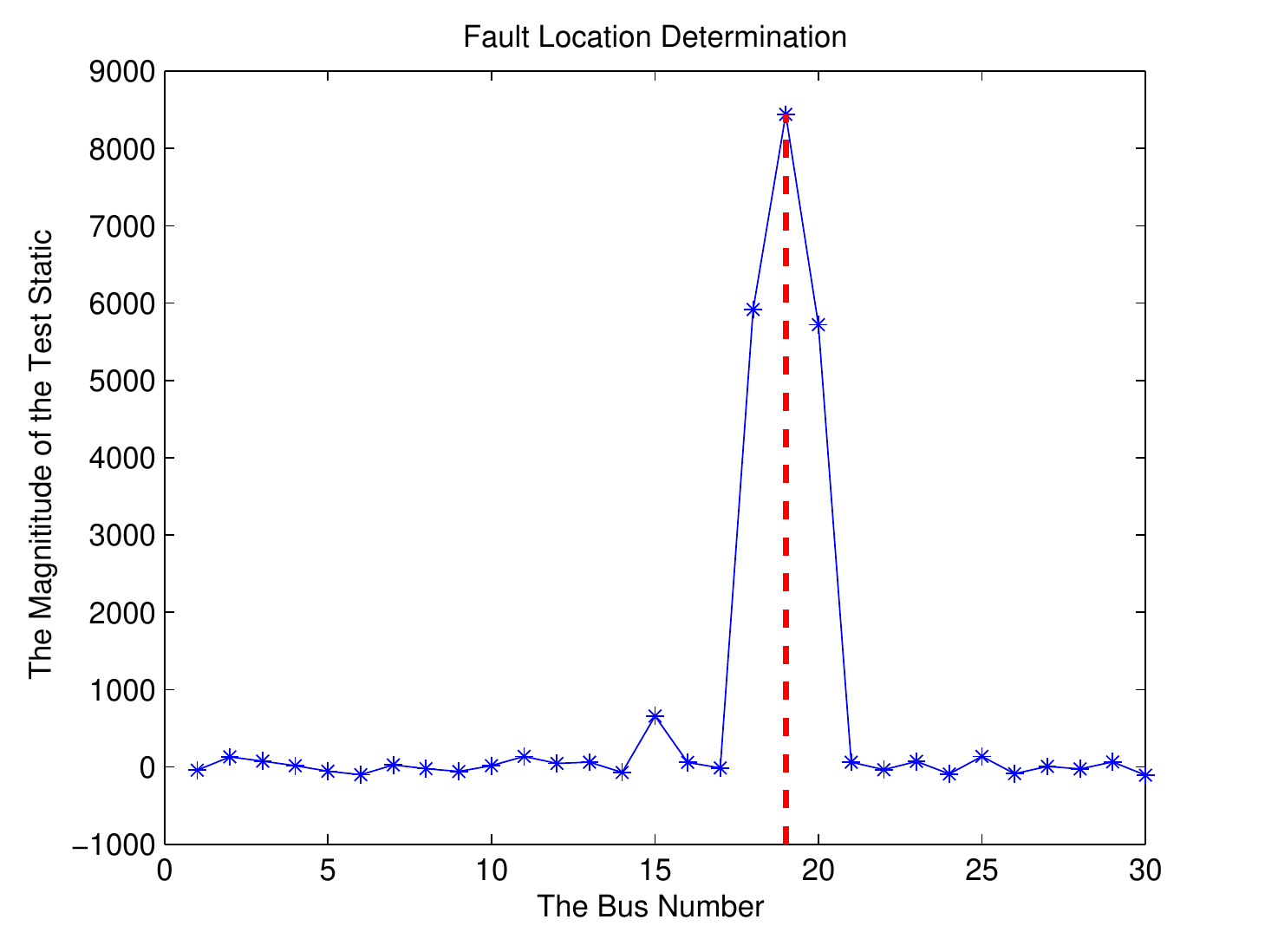}
}
\subfloat[]{\label{fig30gs2}
\includegraphics[width=0.5\columnwidth]{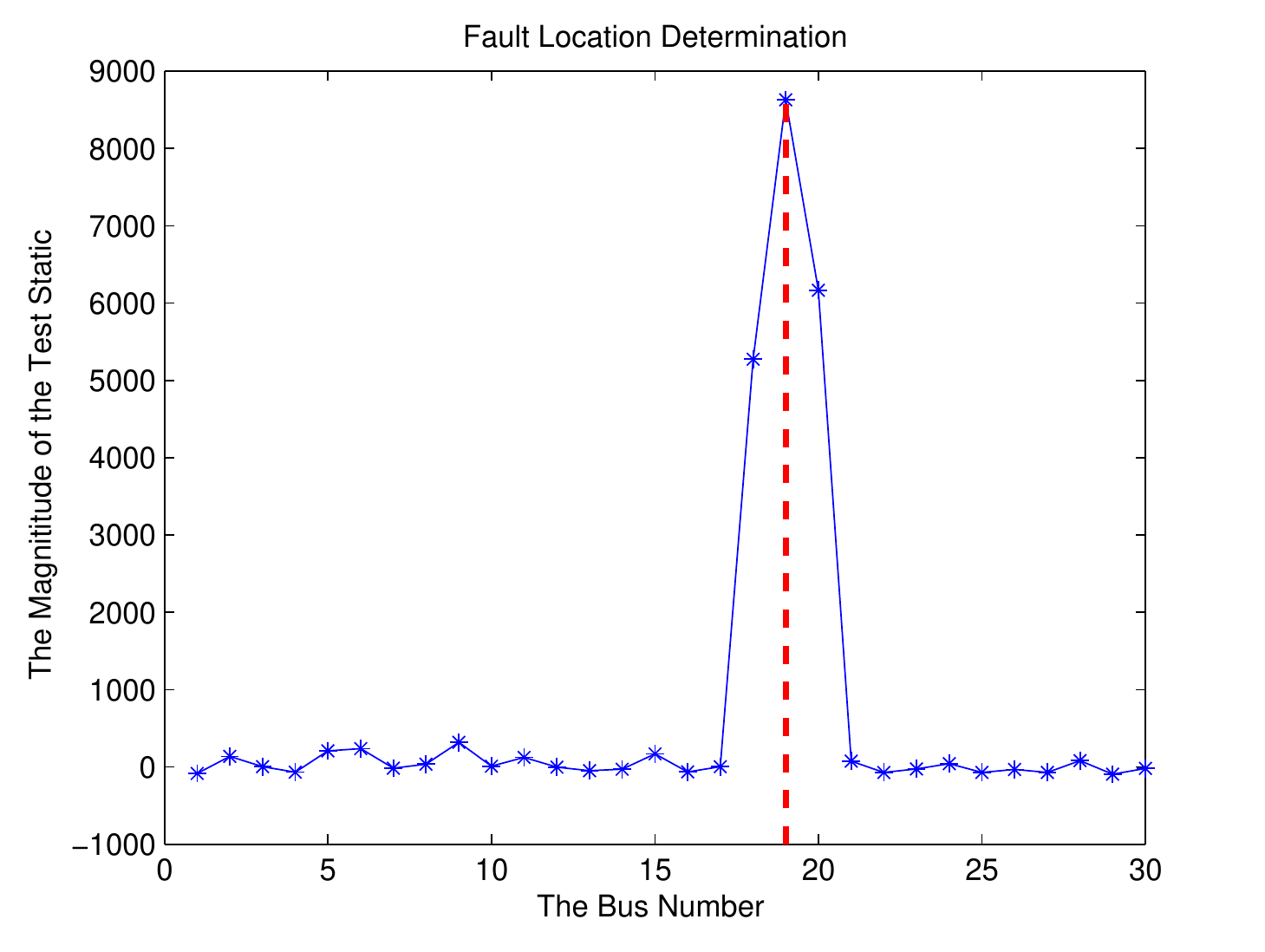}
}\\
\subfloat[]{\label{fig30gs3}
\includegraphics[width=0.5\columnwidth]{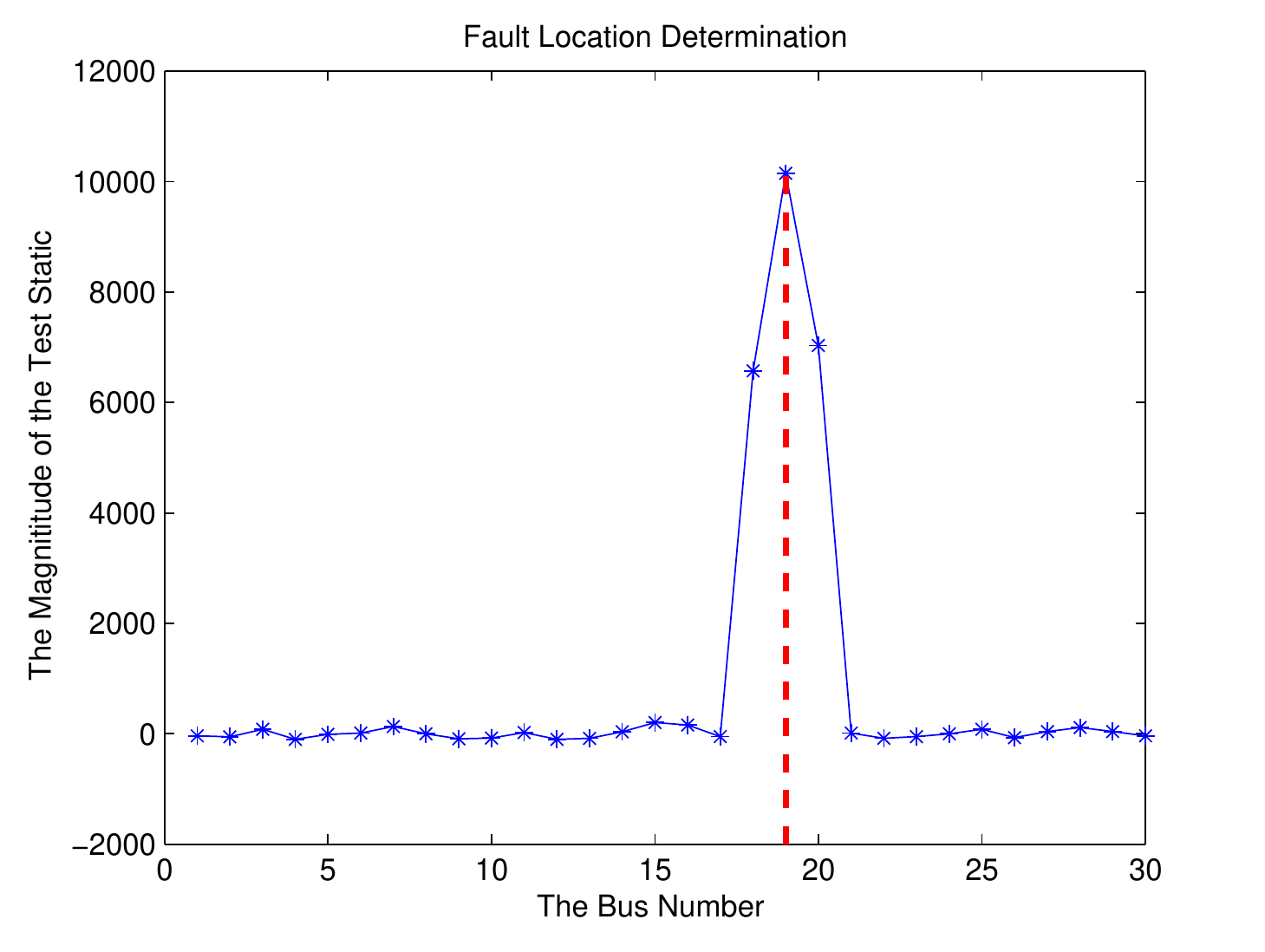}
}
\subfloat[]{\label{fig30gm1}
\includegraphics[width=0.5\columnwidth]{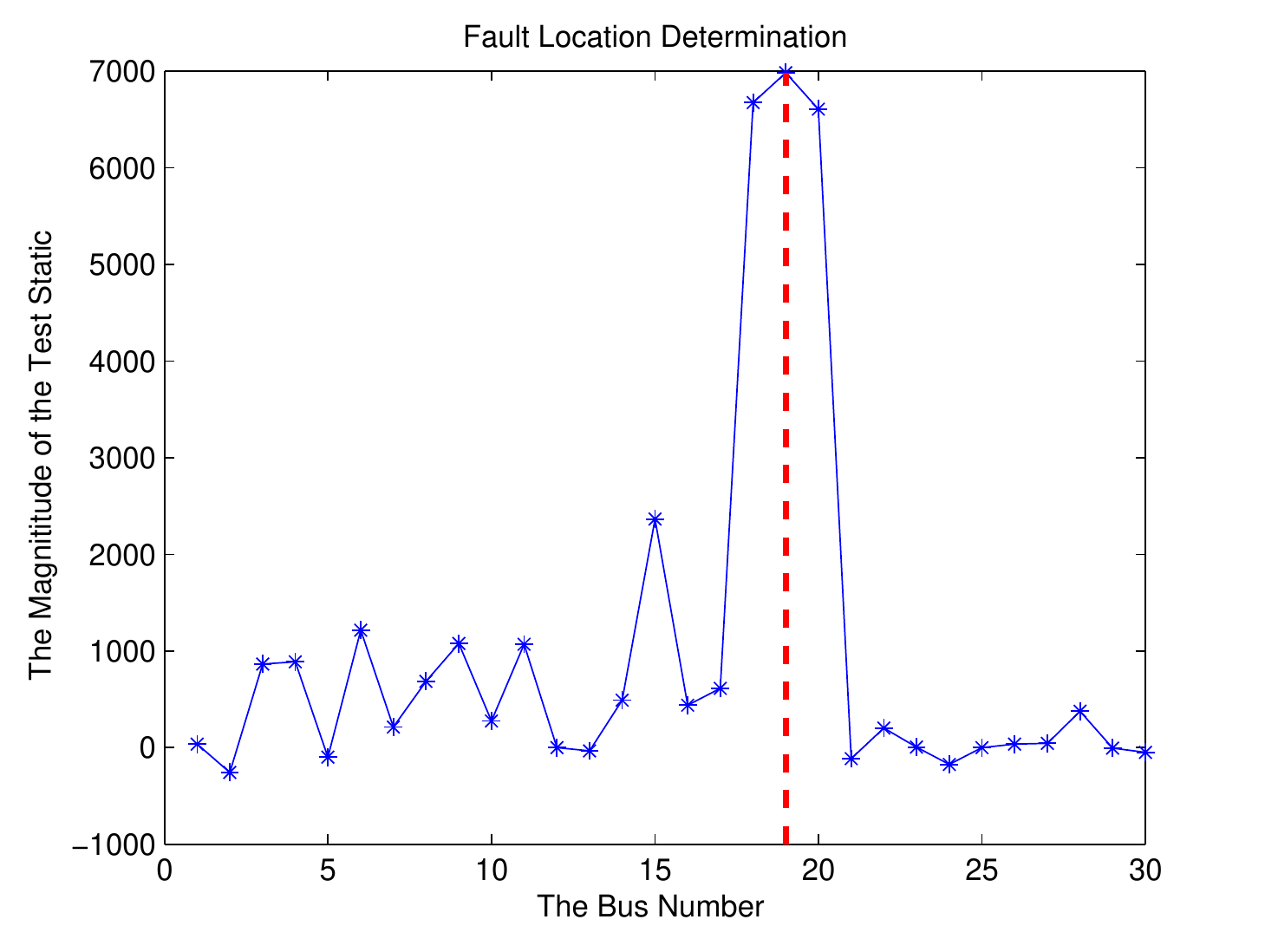}
}\\
\subfloat[]{\label{fig30gm2}
\includegraphics[width=0.5\columnwidth]{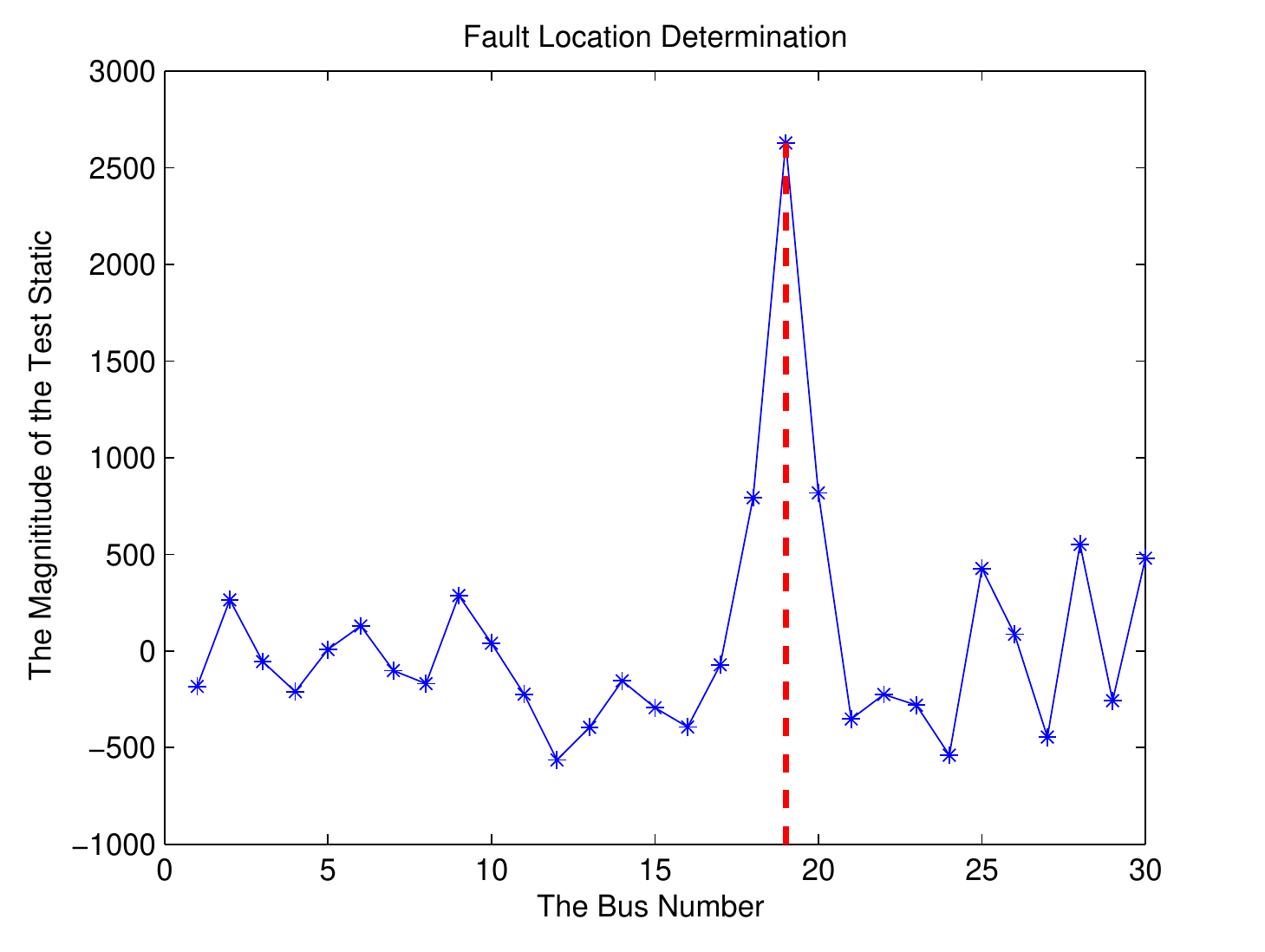}
}
\subfloat[]{\label{fig30gm3}
\includegraphics[width=0.5\columnwidth]{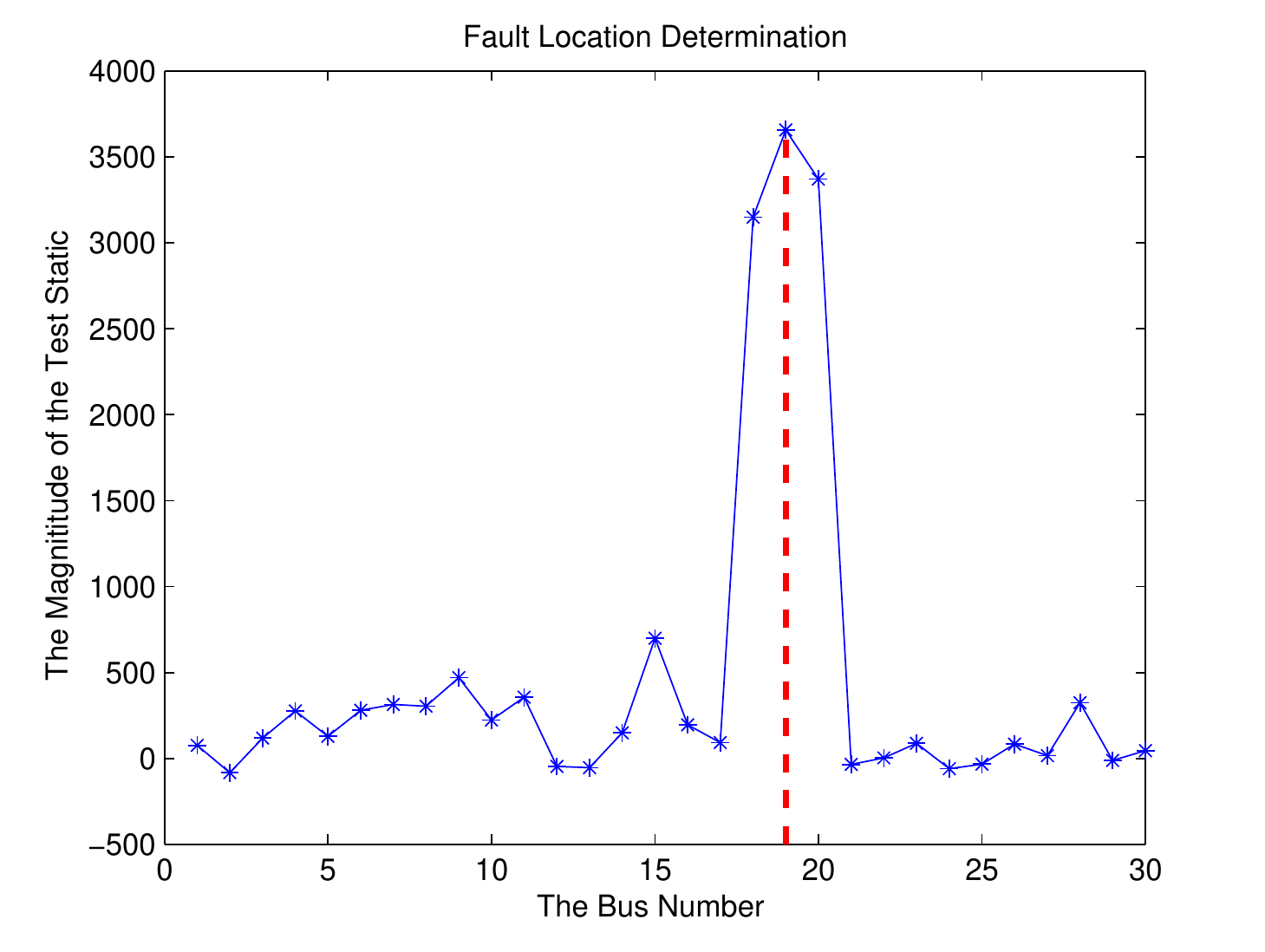}
}
\caption{Determination of most sensitive bus for IEEE 30-bus system. The system were effected by Type I signal with GSN, Type I signal with GMN, Type II signal with GSN, Type II signal with GNN, Type III signal with GSN and Type III signal with GMN, respectively.}
\label{figcase30}
\end{figure}

\begin{figure}[htbp]
\centering
\subfloat[Parameter learning with GSN.]{ \label{fig2383gauss}
\includegraphics[width=0.5\columnwidth]{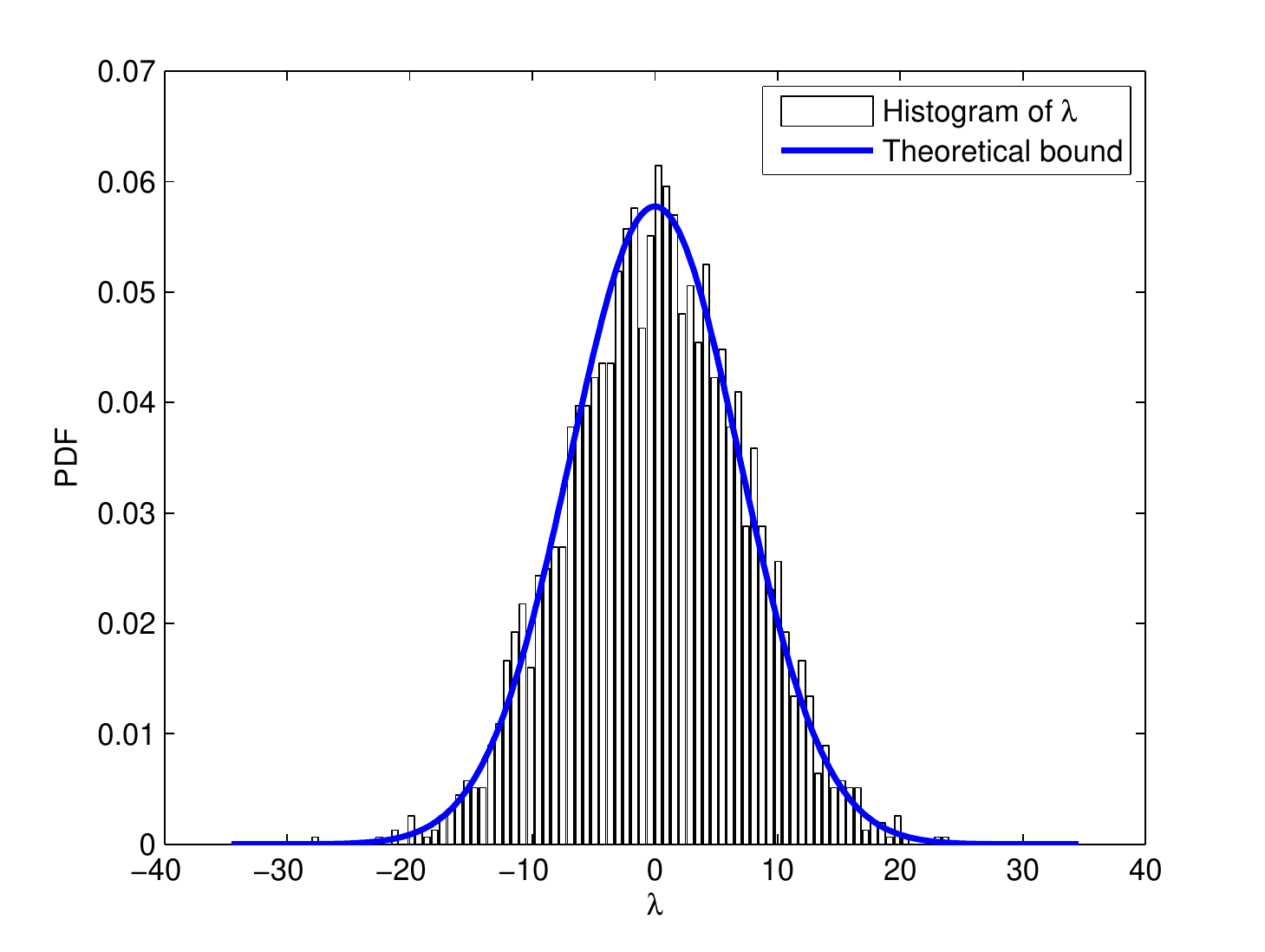}
}
\subfloat[Parameter learning with GMN.]{ \label{fig2383gama}
\includegraphics[width=0.5\columnwidth]{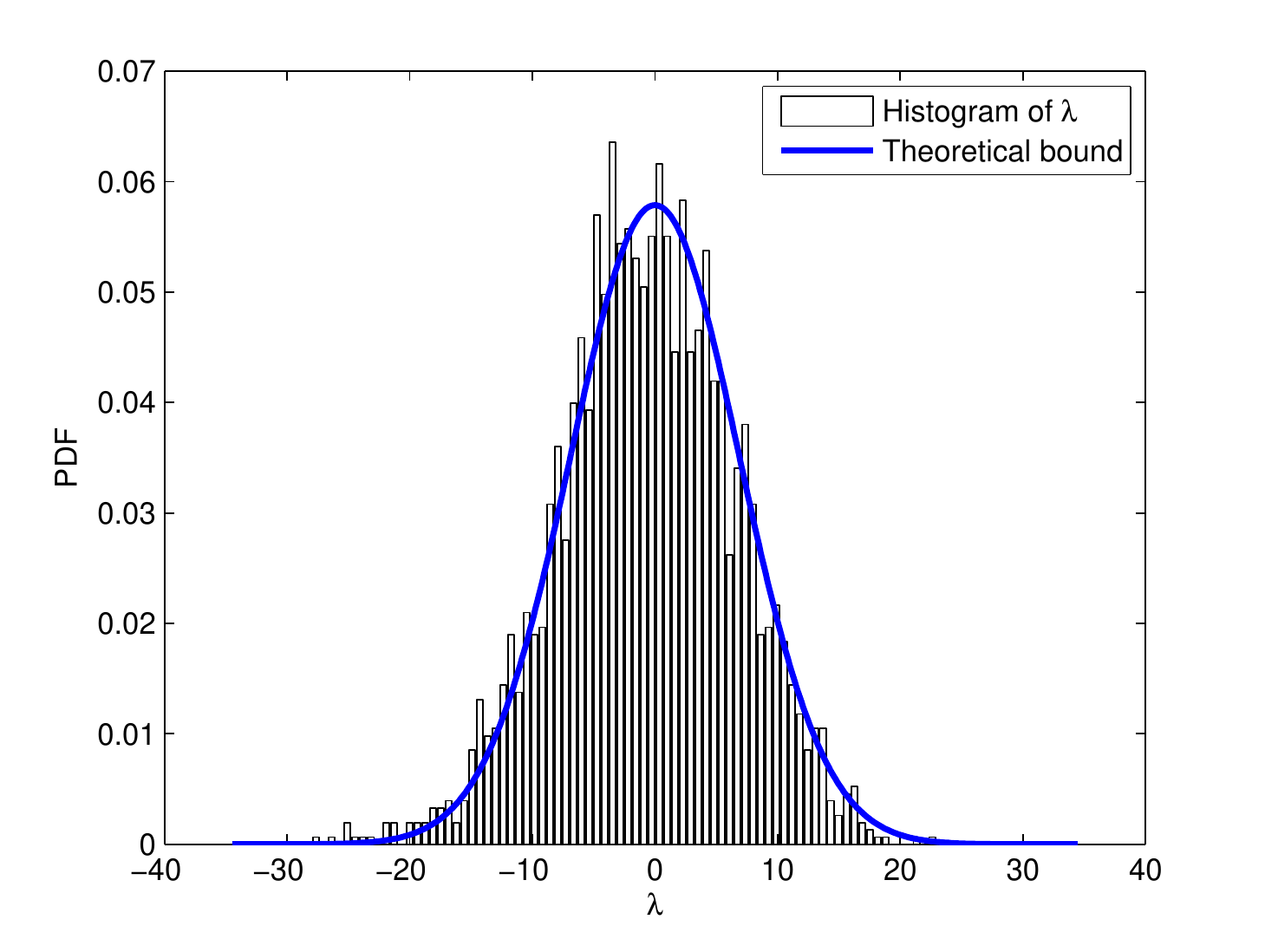}
}
\caption{Parameter learning of the Polish 2383-bus system}
\label{fig2383learn}
\end{figure}

\begin{figure}[htbp]
\subfloat[]{\label{fig2383gs1}
\includegraphics[width=0.5\columnwidth]{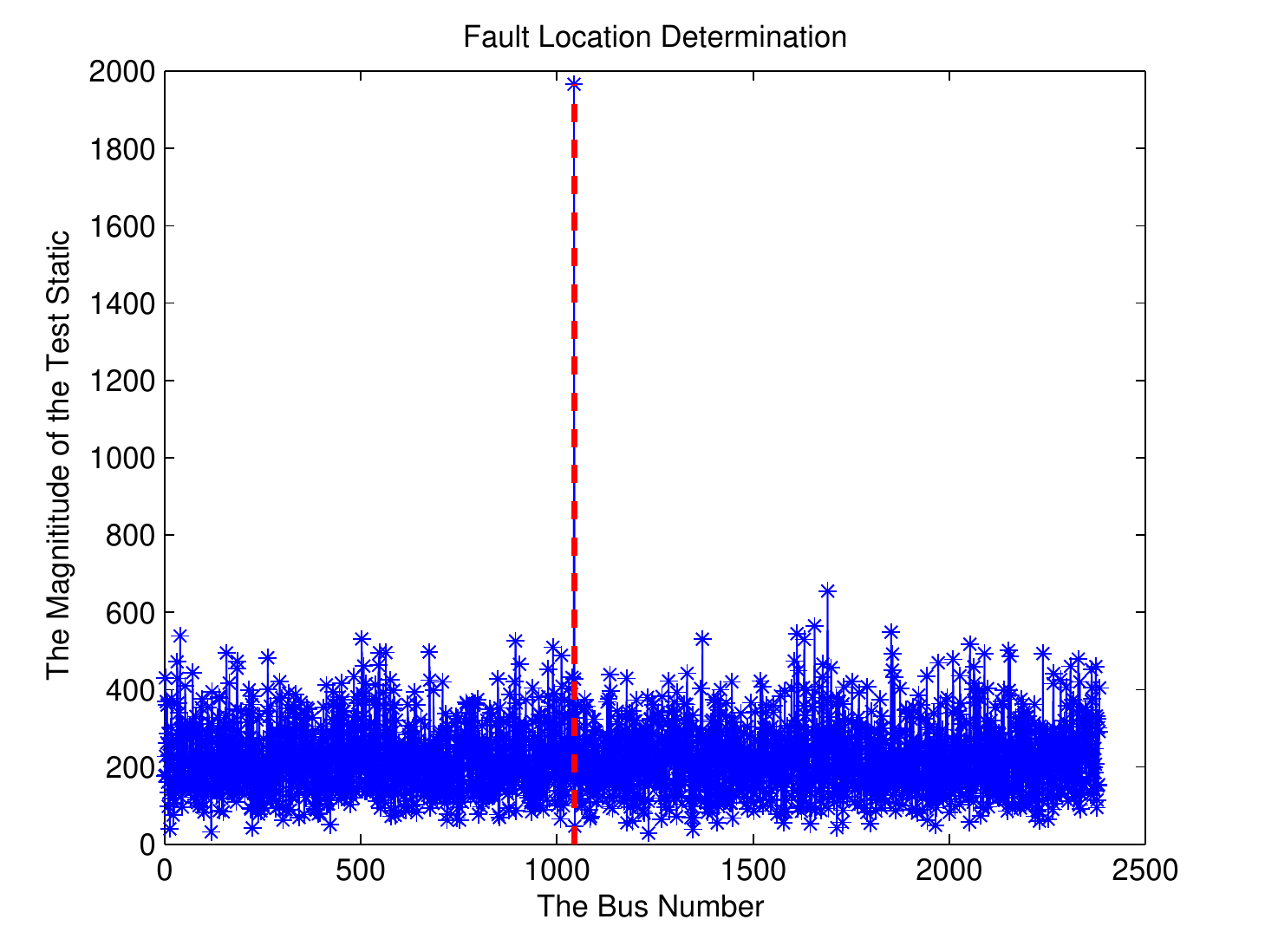}
}
\subfloat[]{\label{fig2383gs2}
\includegraphics[width=0.5\columnwidth]{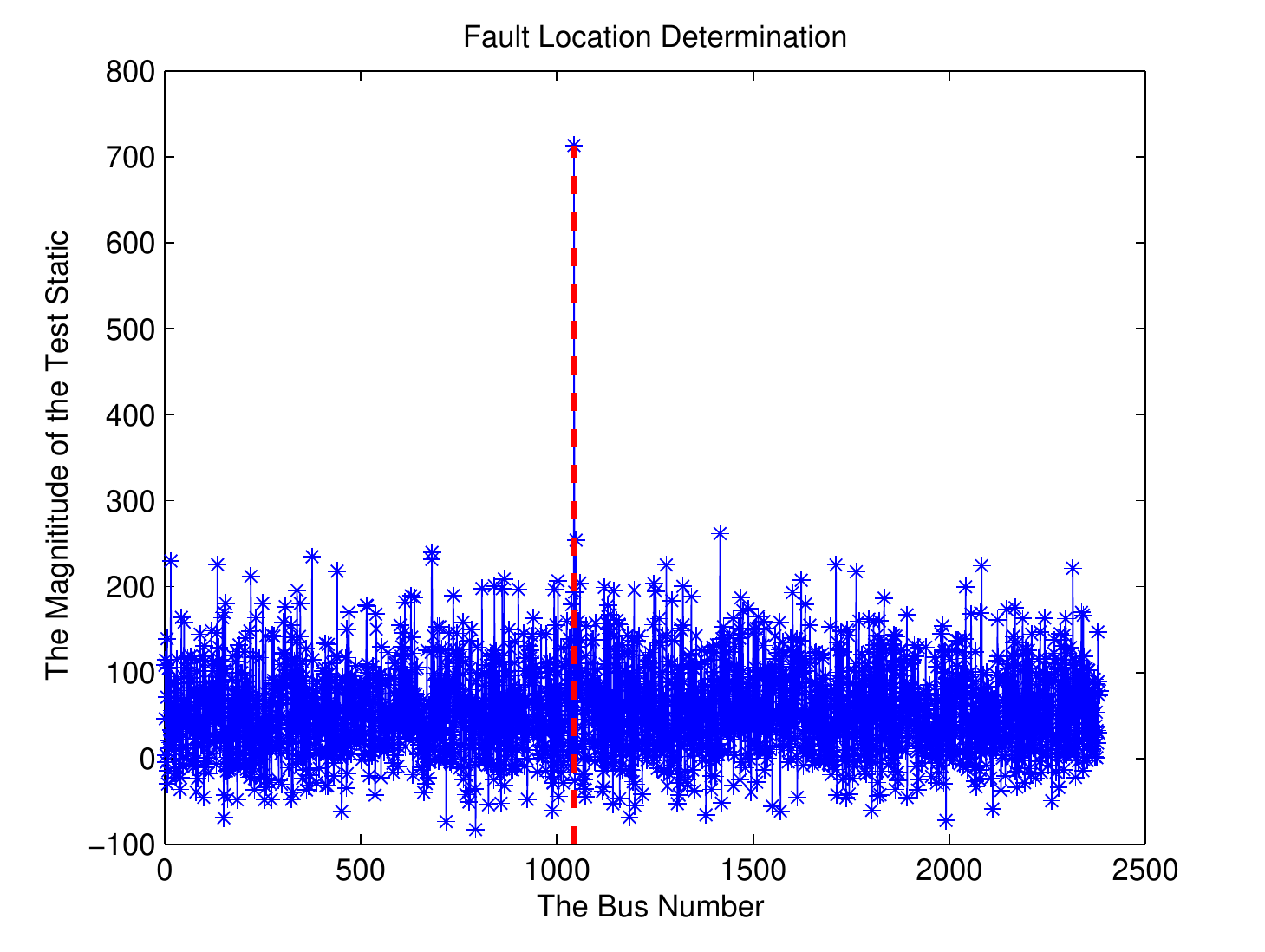}
}\\
\subfloat[]{\label{fig2383gs3}
\includegraphics[width=0.5\columnwidth]{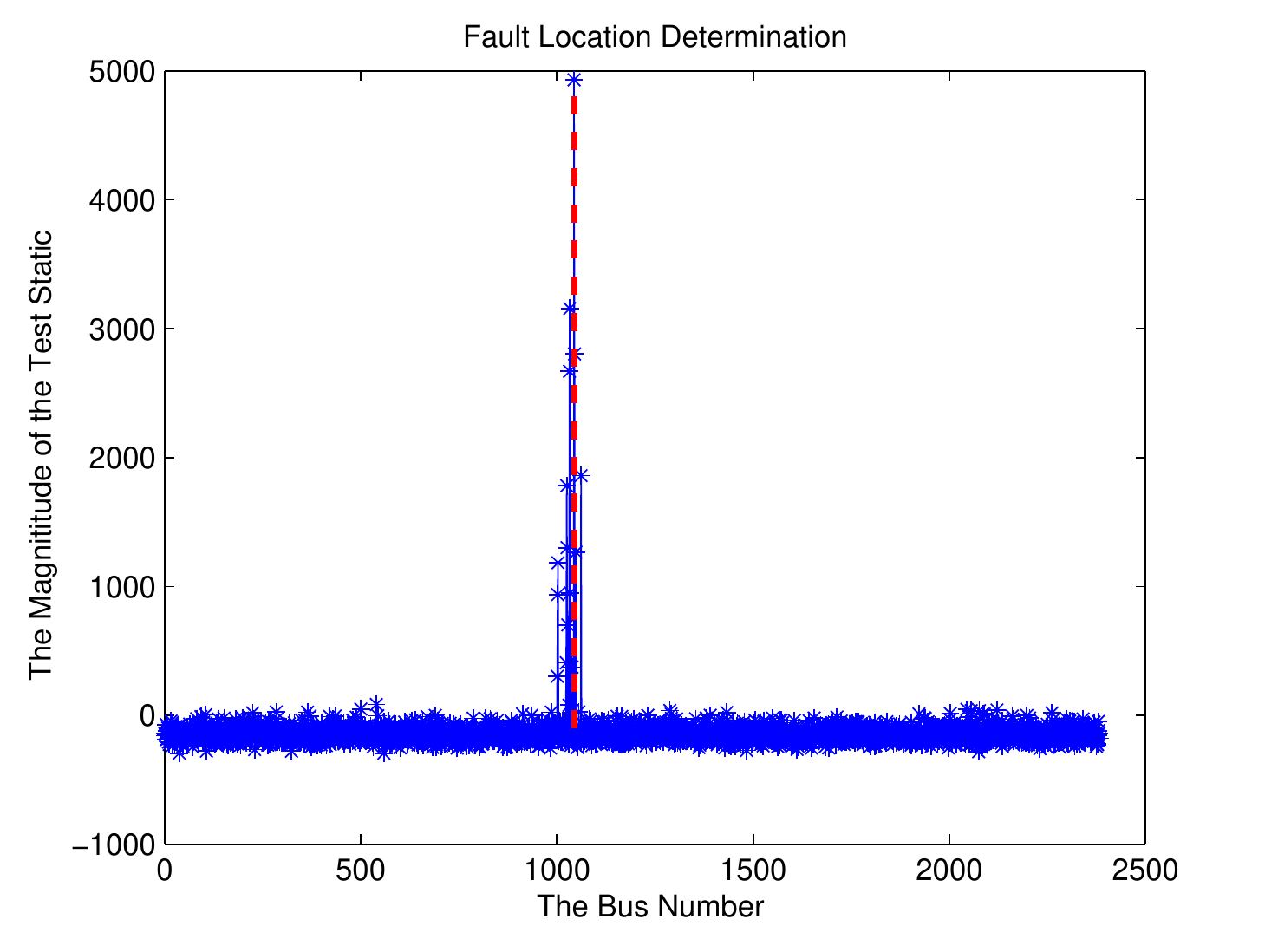}
}
\subfloat[]{\label{fig2383gm1}
\includegraphics[width=0.5\columnwidth]{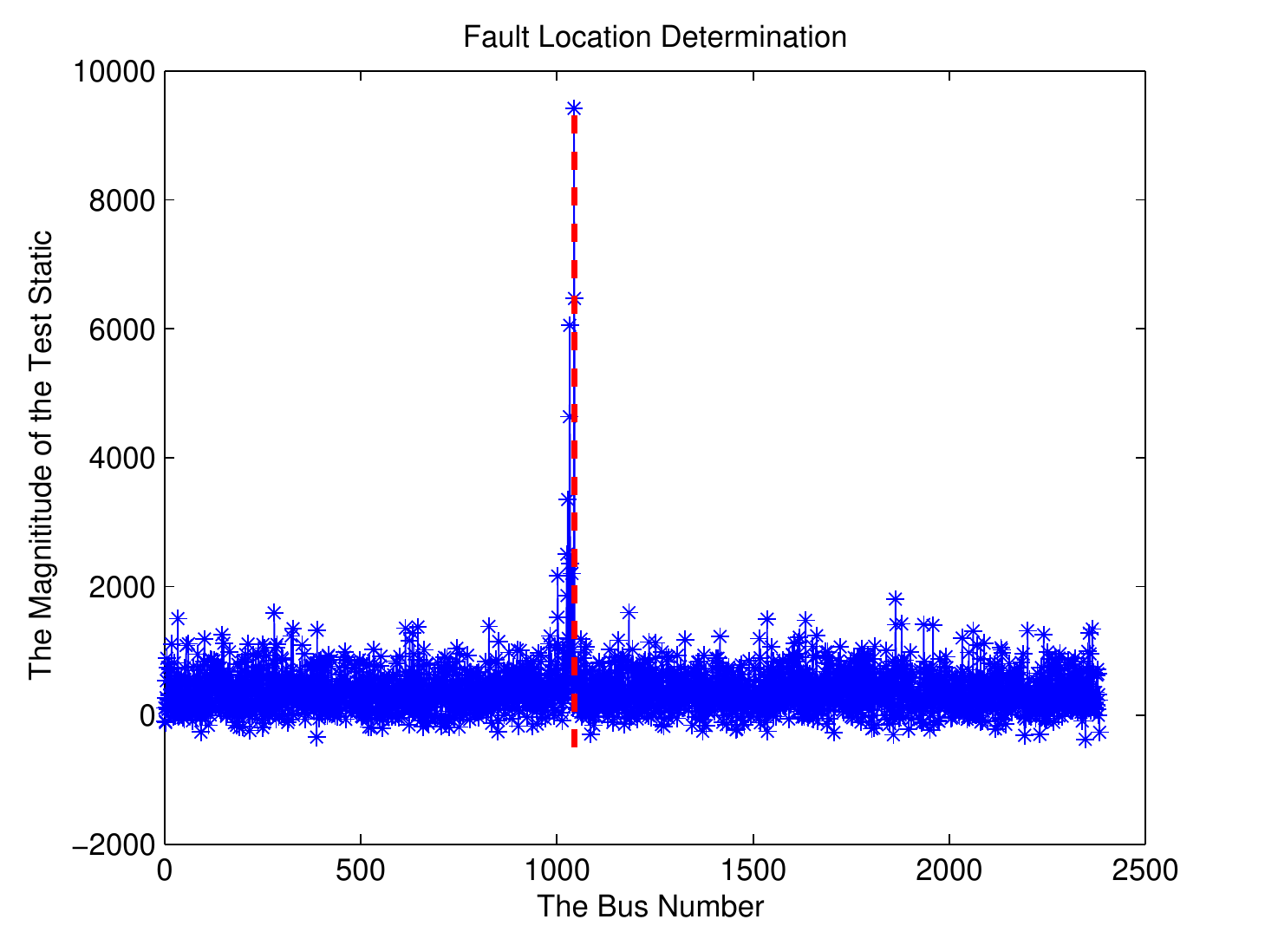}
}\\
\subfloat[]{\label{fig2383gm2}
\includegraphics[width=0.5\columnwidth]{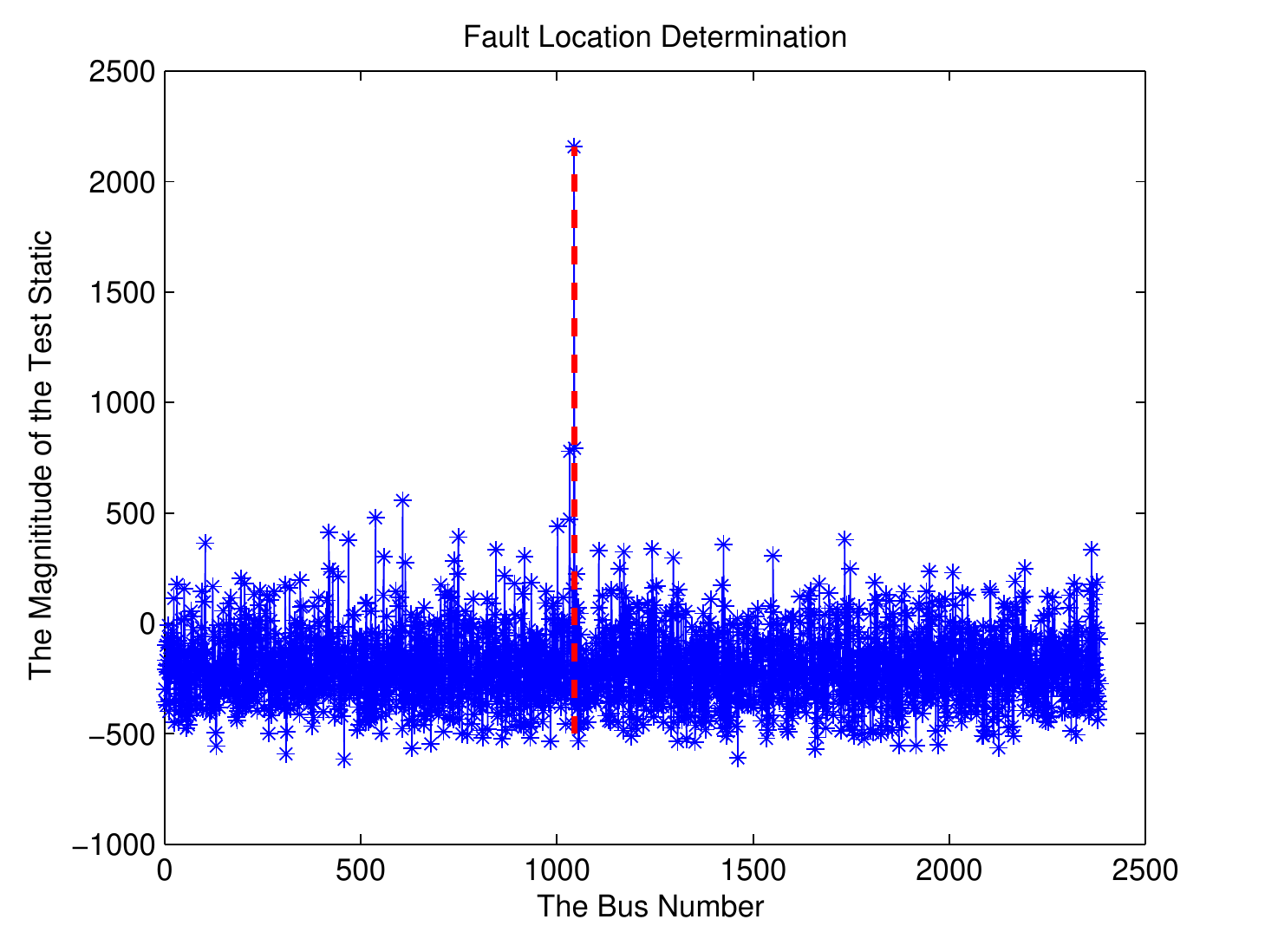}
}
\subfloat[]{\label{fig2383gm3}
\includegraphics[width=0.5\columnwidth]{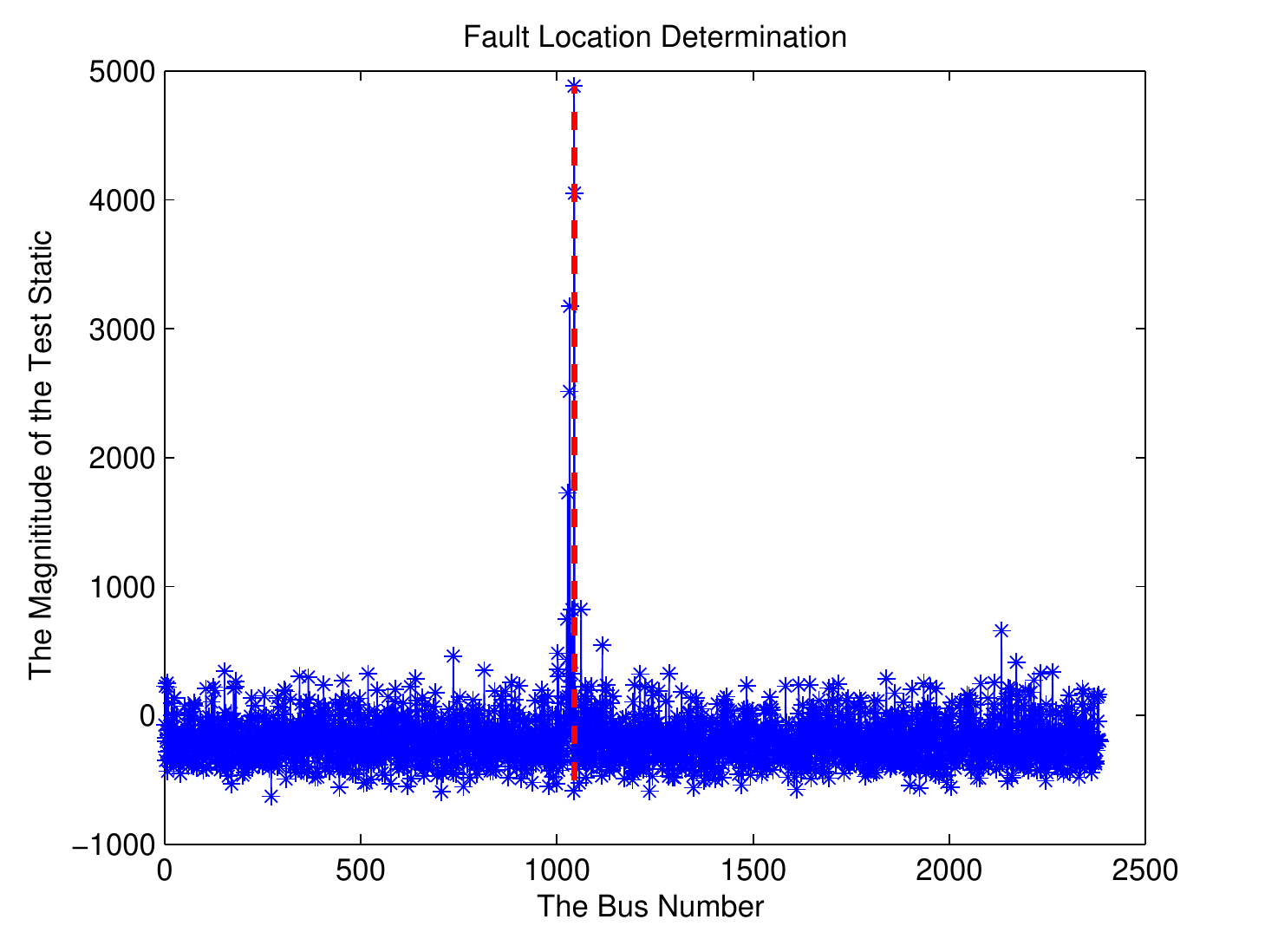}
}
\caption{Determination of most sensitive bus for the Polish 2383-bus system. (The system were effected by Type I signal with GSN, Type I signal with GMN, Type II signal with GSN, Type II signal with GNN, Type III signal with GSN and Type III signal with GMN, respectively.)}
\label{figcase2383}
\end{figure}

The signals were generated in load of P-V node $\rho = 19$ and $\rho = 1044$ for the case of IEEE 30-bus and the Polish 2383-bus system, respectively. Other experimental conditions were the same as the tests for the IEEE 118-bus system. The experiment results are shown in Fig.\ref{fig30learn}, Fig.\ref{figcase30}, Fig.\ref{fig2383learn} and Fig.\ref{figcase2383}.

%

\end{appendices}

\bibliographystyle{IEEEtran}
\bibliography{bib}

\begin{IEEEbiography}
{Lei Chu}
has been pursuing the Ph.D degree at Shanghai Jiaotong University, since 2015. He is also a research assistant in the Big Data Engineering Technology and Research Center, Shanghai. He has published two book chapters, several journal papers and patents. His research interests are in the theoretical and algorithmic studies in signal processing, statistical learning for high dimensional data, random matrix theory, deep learning, as well as their applications in wireless communications, bioengineering and smart grid.
\end{IEEEbiography}

\begin{IEEEbiography}
{Robert Qiu}
(IEEE S'93-M'96-SM'01-FM¡¯14) received the Ph.D. degree in electrical engineering from New York University (former Polytechnic University, Brooklyn, NY). He  is currently a Professor in the Department of Electrical and Computer Engineering, Center for Manufacturing Research, Tennessee Technological University, Cookeville, Tennessee, where he started as an Associate Professor in 2003 before he became a Professor in 2008.  He has also been with the Department of Electrical Engineering, Research Center for Big Data Engineering and Technologies, State Energy Smart Grid R$\&$D Center, Shanghai Jiaotong University since 2015.  His current interest is in wireless communication and networking, machine learning and the Smart Grid technologies. He was Founder-CEO and President of Wiscom Technologies, Inc., manufacturing and marketing WCDMA chipsets. Wiscom was sold to Intel in 2003. Prior to Wiscom, he worked for GTE Labs, Inc. (now Verizon), Waltham, MA, and Bell Labs, Lucent, Whippany, NJ. He has worked in wireless communications and network, machine learning, Smart Grid, digital signal processing, EM scattering, composite absorbing materials, RF microelectronics, UWB, underwater acoustics, and fiber optics. He holds over 6 patents and authored over 70 journal papers/book chapters and 90 conference papers. He has 15 contributions to 3GPP and IEEE standards bodies. In 1998 he developed the first three courses on 3G for Bell Labs researchers. He served as an adjunct professor in Polytechnic University, Brooklyn, New York. Dr. Qiu serves as Associate Editor, IEEE TRANSACTIONS ON VEHICULAR TECHNOLOGY and other international journals. He is a co-author of Cognitive Radio Communication and Networking: Principles and Practice (John Wiley), 2012 and Cognitive Networked Sensing: A Big Data Way (Springer), 2013, and the author of Big Data and Smart Grid (John Wiley), 2015. He is a Guest Book Editor for Ultra-Wideband (UWB) Wireless Communications (New York: Wiley, 2005), and three special issues on UWB including the IEEE JOURNAL ON SELECTED AREAS IN COMMUNICATIONS, IEEE TRANSACTIONS ON VEHICULAR TECHNOLOLOGY and IEEE TRANSACTION ON SMART GRID. He serves as a Member of TPC for GLOBECOM, ICC, WCNC, MILCOM, ICUWB, etc. In addition, he served on the advisory board of the New Jersey Center for Wireless Telecommunications (NJCWT). He is included in Marquis Who¡¯s Who in America.
\end{IEEEbiography}
\begin{IEEEbiography}
{Xing He}
received the bachelor¡¯s and master¡¯s degree in electrical engineering from Southeast University, Nanjing and Shanghai Jiaotong University, Shanghai, in 2008 and 2012, respectively. Currently he is pursuing his Ph.D. studies at School of Electronic Information and Electrical Engineering, Shanghai Jiaotong University, Shanghai, China. His research interests include big data modeling and anomaly detection.
\end{IEEEbiography}

\begin{IEEEbiography}
{Zenan Ling}
received the bachelor¡¯s degree in Department of Mathematics from Nanjing University in 2015. He has been pursuing the Ph.D degree at Shanghai Jiaotong University, since 2015. He is also a research assistant in the Big Data Engineering Technology and Research Center, Shanghai. His research interests are in the machine learning, random matrix and free probability as well as their application in smart grid, stock market and image processing.
\end{IEEEbiography}
\begin{IEEEbiography}
{Yadong Liu }
 was born in Wuhan, China. He received his M.D and Ph.D. of electrical engineering from Shanghai Jiao Tong University, Shanghai, China, in 2008 and 2012 respectively. Currently, he is a lecturer in school of electronic information and electrical engineering at SJTU. His research interests include detecting fault location and diagnosis of power equipment..
\end{IEEEbiography}

\end{document}